\documentclass[10pt]{article}
\usepackage{bbm}
\usepackage[final]{graphics}
\usepackage{amsmath}
\usepackage{amsfonts,amsbsy}
\usepackage{amssymb}
\usepackage{color}
\usepackage{xcolor}
\definecolor{linkcolor}{rgb}{0.4,0.1,0.1}
\definecolor{bibcolor}{rgb}{0.4,0.1,0.1}
\usepackage[colorlinks,linkcolor=linkcolor,citecolor=bibcolor,pdfpagelabels,hyperindex=true]{hyperref}

\def\empile#1\over#2{\mathrel{\mathop{\kern 0pt#1}\limits_{#2}}}
\def\bs{\boldsymbol}
\def\wt#1{\widetilde{#1}}

\newcommand{\slv}{\raise.15ex\hbox{$/$}\kern-.53em\hbox{$v$}}
\newcommand{\slF}{\raise.15ex\hbox{$/$}\kern-.53em\hbox{$F$}}
\newcommand{\slL}{\raise.15ex\hbox{$/$}\kern-.53em\hbox{$L$}}
\newcommand{\slP}{\raise.15ex\hbox{$/$}\kern-.53em\hbox{$P$}}
\newcommand{\slp}{\raise.15ex\hbox{$/$}\kern-.53em\hbox{$p$}}
\newcommand{\slq}{\raise.15ex\hbox{$/$}\kern-.53em\hbox{$q$}}
\newcommand{\slR}{\raise.15ex\hbox{$/$}\kern-.53em\hbox{$R$}}
\newcommand{\slQ}{\raise.15ex\hbox{$/$}\kern-.53em\hbox{$Q$}}
\newcommand{\slK}{\raise.15ex\hbox{$/$}\kern-.53em\hbox{$K$}}
\newcommand{\slk}{\raise.15ex\hbox{$/$}\kern-.53em\hbox{$k$}}
\newcommand{\slD}{\raise.15ex\hbox{$/$}\kern-.73em\hbox{$D$}}
\newcommand{\slC}{\raise.15ex\hbox{$/$}\kern-.53em\hbox{$C$}}
\newcommand{\slA}{\raise.15ex\hbox{$/$}\kern-.53em\hbox{$A$}}
\newcommand{\slSigma}{\raise.15ex\hbox{$/$}\kern-.53em\hbox{$\Sigma$}}
\newcommand{\slpartial}{\raise.15ex\hbox{$/$}\kern-.53em\hbox{$\partial$}}
\newcommand{\slcalP}{\raise.15ex\hbox{$/$}\kern-.63em\hbox{$\cal P$}}

\def\p{{\boldsymbol p}}

\def\k{{\boldsymbol k}}

\def\x{{\boldsymbol x}}

\def\v{{\boldsymbol v}}

\def\rmd{{\rm d}}

\begin{document}

\title{\bf %
Properties of the Boltzmann equation\\ in the classical approximation}
\author{Thomas Epelbaum${}^{1}$, Fran\c cois Gelis${}^{1}$, Naoto Tanji${}^{2,3}$, Bin Wu${}^{1}$}
\maketitle
 \begin{center}
   \begin{enumerate}
   \item Institut de Physique Th\'eorique (URA 2306 du CNRS)\\
     CEA/DSM/Saclay, 91191 Gif-sur-Yvette Cedex, France
   \item Theoretical Research Division, Nishina Center, RIKEN\\
     Wako 351-0198, Japan
   \item Physics Department, Brookhaven National Laboratory\\
     Upton, NY 11973, USA
   \end{enumerate}
 \end{center}

\begin{abstract}
  We study the Boltzmann equation with elastic point-like scalar
  interactions in {two different versions of the} the classical
  approximation. Although solving numerically the Boltzmann equation
  with the unapproximated collision term poses no problem, this allows
  one to study the effect of the ultraviolet cutoff in these
  approximations. This cutoff dependence in the classical
  approximations of the Boltzmann equation is closely related to
  the non-renormalizability of the classical statistical approximation
  of the underlying quantum field theory. The kinetic theory setup
  that we consider here allows one to study in a much simpler way the
  dependence on the ultraviolet cutoff, since one has also access to
  the non-approximated result for comparison.
\end{abstract}

\section{Introduction}
In the early stages of high energy heavy ion collisions, a dense
gluonic matter is formed, nicknamed ``glasma''~\cite{LappiM1}. At
leading order in the strong coupling constant $g^2$, the evolution of
these gluon fields is very simple, since it obeys the classical
Yang-Mills equations~\cite{GelisIJV1}. However, at this order, it does
not thermalize, nor does it exhibit the quasi-perfect hydrodynamical
behavior that seems to be required by many bulk observables in RHIC
and LHC data~\cite{FukusG1}. Next-to-leading order corrections to
observables are quadratic in small perturbations around the LO
classical gauge field~\cite{GelisV2,GelisLV3}. However, due to the
Weibel instability, these perturbations grow exponentially in
time~\cite{RomatV3}, and the NLO corrections can become larger than
the LO. The resummation of an infinite class of higher order
contributions, at all loop orders, is necessary in order to stabilize
the perturbative expansion.

The Classical Statistical Approximation (CSA) is one of the schemes
that has been employed recently in order to realize such a
resummation~\cite{DusliEGV1,EpelbG1,EpelbG3,BergeBS1,BergeSS3,BergeBSV1,BergeBSV2,BergeBSV3}. Since
it amounts to averaging solutions of the classical equations of motion
over a Gaussian ensemble of initial conditions, its practical
implementation is not more difficult than the LO calculation
itself. Moreover, it trivially gives gauge invariant results for
physical observables.  By adjusting appropriately the 2-point
correlation function that characterizes the Gaussian fluctuations of
the initial classical fields, one can set up the CSA in such a way
that it reproduces exactly the LO and NLO results, plus a subset of
the higher order contributions~\cite{FukusGM1,DusliGV1,EpelbG2}. By
construction, the CSA is affected by the ultraviolet divergences that
arise in loop corrections. For instance, all the divergences present
at 1-loop are also present in the CSA since it is built precisely to
reproduce exactly the NLO result. Practical implementations of the CSA
represent space on a lattice, which provides a natural ultraviolet
cutoff\footnote{The loop momentum that causes the UV divergences is
  the momentum of the initial fluctuations. On the lattice, this
  momentum cannot exceed the inverse lattice spacing. However, it is
  also possible to limit by hand the largest momentum mode included in
  these initial fluctuations. When doing this, the effective UV
  regulator is no longer the inverse lattice spacing, but this cutoff
  that has been introduced in the  spectrum of fluctuations.}.

It was observed in large scale lattice simulations using the CSA
\cite{BergeBSV3} that this approximation produces results that depend
strongly on the ultraviolet cutoff when it is chosen to be much larger
than the physical scales of the problem. In ref.~\cite{EpelbGW1}, it
was shown that the CSA is a non-renormalizable approximation of the
underlying quantum field theory. The problem starts at 2-loops (the
order where the CSA first differs from the underlying theory), and is
due to the fact that the CSA includes some quantum corrections but not
all of them\footnote{The CSA captures correctly the quantum
  corrections to the initial state, while the time evolution remains
  purely classical. Renormalizability requires that both types of
  quantum corrections be treated on the same footing.}.  Among the
Feynman diagrams that contain non-renormalizable contributions, one
can find some 2-loop self-energies that enter in the collision term of
the Boltzmann equation, in the standard derivation of kinetic theory
from the Dyson-Schwinger equations. This suggests that a similar
cutoff dependence should be found in the classical approximation of
the Boltzmann equation, and that one could investigate the ultraviolet
artifacts of the CSA in the much simpler setting of kinetic
theory. This would be particularly useful because the Boltzmann
equation with $2\to 2$ elastic scatterings is simple enough to be
solved numerically without doing the classical approximation, thereby
provided the ``exact'' result as a reference, something that cannot be
done in the underlying quantum field theory\footnote{{A similar
    kinetic treatment, although looking at a different problem, was
    recently performed in Yang-Mills theory \cite{YorkKLM1,KurkeL1},
    using the Boltzmann equation derived in \cite{ArnolMY5}}.}.

In this paper, we start from the classical approximation of the
Boltzmann equation already discussed in refs.~\cite{MuellS1,Jeon3},
and we focus on issues related to the dependence of its solutions on
the ultraviolet cutoff. In fact, we consider two versions of the
classical approximation, that correspond to including or not
vacuum fluctuations in the underlying quantum field theory. We will
compare the solutions of the unapproximated Boltzmann equation with
the solutions obtained in these two approximations.

Our paper is organized as follows. In the section \ref{sec:approx}, we
rederive the classical approximations of the Boltzmann equation, by
following the same approach as in ref.~\cite{EpelbGW1} in order to
emphasize the fact that the approximation is identical to the one that
leads to the CSA.  In the section \ref{sec:bec}, we generalize the
Boltzmann equation in order to include the possibility of
Bose-Einstein condensation. This is necessary if one wants to consider
overoccupied initial conditions, as would be the matter formed in
heavy ion collisions. In the section \ref{sec:UVcoll}, we first study
the explicit dependence on the UV cutoff of the collision term in the
classical approximation. In one version of this approximation, the
collision term contains a piece which is quadratic in the UV cutoff,
and is a direct consequence of the non-renormalizable contributions
exhibited in ref.~\cite{EpelbGW1}. In the section \ref{sec:asympt}, we
study the fixed points of the Boltzmann equation in the two versions
of the classical approximation, by extending to these modified
equations the standard proof of the H-theorem. We also show that the
fixed points in the classical approximation are strongly dependent on
the UV cutoff, in a way which is in perfect agreement with the CSA
computations of ref.~\cite{BergeBSV3}. Finally, in the section
\ref{sec:evol}, we compare the time evolutions in the classical
approximations and in the unapproximated Boltzmann equation.  The
appendices contain some more technical material. The appendix
\ref{app:sigma} gives some details bout the calculation of the
self-energies that enter in the expression of the collision term. In
the appendix \ref{app:coll}, we remind the reader about the reduction
of the collision term to a 2-dimensional integral (for an isotropic
particle distribution), and we give some details about the
discretization we use in order to compute it. In the appendix
\ref{app:lin}, we give explicit formulas for the spurious linear terms
that appear in the classical collision term. { In the appendix
  \ref{app:masslessnc}, we give a brief discussion about how the
  existence of a Bose-Einstein condensate depends on the UV cutoff.}

\section{Boltzmann eq. in the classical approximation}
\label{sec:approx}
\subsection{Collision term in the Schwinger-Keldysh formalism}
Our starting point will be the Boltzmann equation written in the
following form,
\begin{equation}
[\partial_t+\v_\p\cdot{\bs\nabla}]\,f(p)=\underbrace{\frac{i}{2\omega_\p}
\left[f(\p)\Sigma_{-+}(P)-(1+f(\p))\Sigma_{+-}(P)\right]}_{{\cal C}_\p[f]}\; ,
\label{eq:Boltz-plain}
\end{equation}
where $\Sigma_{+-}$ and $\Sigma_{-+}$ are the self-energies in the
Schwinger-Keldysh formalism~\cite{Schwi1,Keldy1}, evaluated for an
on-shell momentum $P$.  Although $f(\p)$ is spacetime dependent, we do
not write explicitly the time and space arguments in order to lighten
the notations. The self-energies $\Sigma_{+-}$ and $\Sigma_{-+}$ that
enter in the collision term can be calculated from the perturbative
diagrammatic rules of the Schwinger-Keldysh formalism, in which the
propagators read
\begin{eqnarray}
  G_{++}^0(p)&=&\frac{i}{p^2-m^2+i\epsilon}+2\pi f(\p)\delta(p^2-m^2)
  \nonumber\\
G_{--}^0(p)&=&\frac{-i}{p^2-m^2-i\epsilon}+2\pi f(\p)\delta(p^2-m^2)
  \nonumber\\
  G_{+-}^0(p)&=&2\pi (\theta(-p^0)+f(\p))\delta(p^2-m^2)
  \nonumber\\
  G_{-+}^0(p)&=&2\pi (\theta(p^0)+f(\p))\delta(p^2-m^2)\; ,
\end{eqnarray}
and where the vertices of type $+$ read
${\Gamma_{\hspace{-0.05cm}_{++++}}}=-ig^2$ while those of type $-$
are ${\Gamma_{\hspace{-0.05cm}_{----}}}=+ig^2$.  The lowest order
contribution to the collision term is a 2-loop self-energy, that
corresponds to the elastic $2\to 2 $ scattering process. For the
record, let us recall here the $2\to 2$ contribution to the collision
term for scalar particles,
\begin{eqnarray}
C_\p[f]&&=
\smash{\frac{g^4}{4\omega_\p}\int_{\p'\k\k'}}
(2\pi)^4\delta(P+K-P'-K')\nonumber\\
&&
\qquad\qquad\qquad\qquad\times[f(\p')f(\k')(1+f(\p))(1+f(\k))\nonumber\\
&&
\qquad\qquad\qquad\qquad\quad-f(\p)f(\k)(1+f(\p'))(1+f(\k'))]\; ,
\label{eq:Cfull}
\end{eqnarray}
where for the sake of brevity we have denoted
\begin{equation}
\int_\k\equiv\int\frac{\rmd^3\k}{(2\pi)^3 2\omega_\k}
\end{equation}
the invariant phase space integral for an on-shell particle. 

\subsection{Collision term in the retarded-advanced basis}
In order to make the classical approximation more transparent, we
follow the procedure used in refs.~\cite{MuellS1,Jeon3,EpelbGW1}, that
consists in going from the Schwinger-Keldysh basis to the
retarded-advanced basis~\cite{Keldy1,EijckKW1,Jeon4}. We follow the
notations of the section 2.3 of ref.~\cite{EpelbGW1}, where the
transformation matrix is defined as\footnote{Although in
  ref.~\cite{EpelbGW1}, this transformation was introduced for the
  vacuum Feynman rules, the transformation matrix is in fact medium
  independent. It remains unchanged even if the system contains a
  particle distribution $f(\p)$.}
\begin{equation}
\Omega_{\alpha\epsilon}
\equiv
\begin{pmatrix}
1 & -1 \\
1/2 & 1/2 \\
\end{pmatrix}\; ,
\end{equation}
which leads to the following transformed propagators~:
\begin{eqnarray}
{\mathbbm G}_{\alpha\beta}^0\equiv 
\sum_{\epsilon,\epsilon^\prime=\pm}
\Omega_{\alpha\epsilon}\Omega_{\beta\epsilon^\prime}
{G}_{\epsilon\epsilon^\prime}^0=
\begin{pmatrix}
0 & G_{_A}^0= G_{++}^0-G_{-+}^0\\
G_{_R}^0= G_{++}^0-G_{+-}^0& G_{_S}^0=\frac{1}{2}(G_{++}^0+G_{--}^0)\\
\end{pmatrix}\; .
\label{eq:SK-rotation}
\end{eqnarray}
It is easy to check the self-energies  are related
by the following formula,
\begin{equation}
\Sigma_{\epsilon\epsilon'}
=
\Omega_{\alpha\epsilon}\Omega_{\beta\epsilon'}
{\bs\Sigma}_{\alpha\beta}\; ,
\end{equation}
where ${\bs\Sigma}_{\alpha\beta}$ denotes the self-energy in the
retarded-advanced basis. In particular, we have
\begin{eqnarray}
\Sigma_{+-}&=&
-{\bs\Sigma}_{11}
+\frac{1}{2}{\bs\Sigma}_{12}
-\frac{1}{2}{\bs\Sigma}_{21}
\nonumber\\
\Sigma_{-+}&=&
-{\bs\Sigma}_{11}
-\frac{1}{2}{\bs\Sigma}_{12}
+\frac{1}{2}{\bs\Sigma}_{21}\; .
\end{eqnarray}
Therefore, the collision term is given as
follows in terms of the rotated self-energies~:
\begin{equation}
C_\p[f]=
\frac{i}{2\omega_\p}
\left[
{\bs\Sigma}_{11}(P)
+\left(f(\p)+\frac{1}{2}\right)({\bs\Sigma}_{21}(P)-{\bs\Sigma}_{12}(P))
\right]\; .
\label{eq:coll12}
\end{equation}
Note that ${\bs\Sigma}_{11}$ is purely imaginary, as well as the
difference ${\bs\Sigma}_{21}(P)-{\bs\Sigma}_{12}(P)$, so that the
collision term is real, as it should.

In the retarded-advanced basis, the bare propagators are
\begin{eqnarray}
  &&
  G_{21}^0(p) = \frac{i}{(p^0+i\epsilon)^2-\p^2-m^2}\;,\quad
  G_{12}^0(p) = \frac{i}{(p^0-i\epsilon)^2-\p^2-m^2}\;,\nonumber\\
  &&
  G_{22}^0(p) = 2\pi\,\left(\frac{1}{2}+f(\p)\right)\,\delta(p^2-m^2)\;,
  \label{eq:RA:prop}
\end{eqnarray}
and the non-zero vertices are $\Gamma_{1222}=-ig^2$ and $\Gamma_{1112}=-ig^2/4$.

\subsection{Collision term in the classical approximation}
From the collision term expressed in the form of
eq.~(\ref{eq:coll12}), it is easy to perform an approximation which is
the exact analogue of the classical statistical approximation in the
underlying quantum field theory. The Lagrangian of the theory in the
retarded-advanced basis has two kinds of interaction terms,
$\phi_1\phi_2^3$ and $\phi_1^3\phi_2$, {where
  $2\,\phi_2=\phi^++\phi^-$ and $\phi_1=\phi^+-\phi^-$}. The classical
approximation amounts to neglecting the $\phi_1^3\phi_2$ term. This
limits the possible assignments for the indices $1$ and $2$ when
constructing the graphs that contribute to the self-energies
${\bs\Sigma}_{12}$, ${\bs\Sigma}_{21}$ and ${\bs\Sigma}_{11}$. In the
literature, there are in fact two versions of the classical
approximation (that we will refer to as ${\cal C}^0$ and ${\cal C}^1$
thereafter)~:
\begin{itemize}
\item[${\cal C}^0$~:]~{\sl Neglect both the vertex $\Gamma_{1112}$ and the
    $1/2$ in the propagator $G_{22}^0$.} This approximation coincides
  with the unapproximated theory at Leading Order in the coupling $g$
  in the regime of strong fields ($\phi\sim g^{-1}$) and/or large
  occupation numbers ($f(\p)\sim g^{-2}$).
\item[${\cal C}^1$~:]~{\sl Neglect the vertex $\Gamma_{1112}$ but keep the
    $1/2$ in the propagator $G_{22}^0$.} This approximation coincides
  with the unapproximated theory both at LO and NLO in the regime of
  strong fields ($\phi\sim g^{-1}$) and/or large occupation numbers
  ($f(\p)\sim g^{-2}$).
\end{itemize}
The derivation of the collision term in these approximations is
straightforward (see refs.~\cite{MuellS1,Jeon3}) from the Feynman
rules of the retarded-advanced formalism. If one starts from the full
collision term (before doing the classical approximation), one can
obtain the approximated collision terms by the following truncations
and substitutions~:
\begin{itemize}
\item[${\cal C}^0$~:]~Keep only the terms of highest degree in $f$ in
  the collision term. For the $2\to 2$ collision term, these terms are
  cubic in $f$.
\item[${\cal C}^1$~:]~Keep the terms of highest degree in $f$, and then
  substitute
\begin{equation}
f\quad\to\quad f+\frac{1}{2}\; .
\end{equation}
For the elastic collision term, this gives the correct $f^3$ and $f^2$
terms, but also some spurious terms that are linear in $f$. More
generally, for any contribution to the collision term, it can be shown
that this ansatz gives all the terms with the leading and subleading
powers in $f$. 

\end{itemize}
Explicitly, the elastic collision term ($\p\k\leftrightarrow\p'\k'$)
in the ${\cal C}^0$ classical approximation reads
\begin{eqnarray}
&&C_\p^{{\cal C}^0}[f]=\frac{g^4}{4\omega_\p}
\int_{\k}\int_{\p'}\int_{\k'}
(2\pi)^4\delta(P+K-P'-K')\;
\nonumber\\
&&\qquad\times
\big[f(\p')f(\k')(f(\p)+f(\k))-
f(\p)f(\k)(f(\p')+f(\k'))
\big]\; .
\label{eq:Cclass0}
\end{eqnarray}
For the
${\cal C}^1$ classical approximation, the elastic collision term reads
\begin{eqnarray}
&&C_\p^{{\cal C}^1}[f]=\frac{g^4}{4\omega_\p}
\int_{\k}\int_{\p'}\int_{\k'}
(2\pi)^4\delta(P+K-P'-K')\;
\nonumber\\
&&\qquad\times
\big[(f(\p')+\frac{1}{2})(f(\k')+\frac{1}{2})(1+f(\p)+f(\k))
\nonumber\\
&&\qquad-
(f(\p)+\frac{1}{2})(f(\k)+\frac{1}{2})(1+f(\p')+f(\k'))
\big]\; .
\label{eq:Cclass1}
\end{eqnarray}

Our goal in this paper is to investigate the effect of replacing the
unapproximated collision term $C_\p[f]$ by its classical
approximations, given in eqs.~(\ref{eq:Cclass0}) and
(\ref{eq:Cclass1}), in order to gain some insight about the
limitations of the classical statistical approximation, especially
regarding its dependence on the ultraviolet cutoff.

\section{Boltzmann equation with a BEC}
\label{sec:bec}
If one takes into account only $2\to2$ processes, some initial
conditions can have more particles than can be accommodated in a
Bose-Einstein distribution\footnote{See \cite{BlaizLM2,BlaizWY1} for a similar discussion for a dense system of gluons in perturbative QCD.}. At a given temperature, the maximal number
of particles is achieved when the chemical potential is equal to the
mass of the particles. If the initial condition contains more
particles than this value, then the extra particles will form a
Bose-Einstein condensate (BEC) at $\p=0$\footnote{If we assume that
  the system reaches an equilibrium made of a Bose-Einstein
  distribution with a chemical potential $\mu$, plus some extra
  particles of momentum $\p_*$, then it is possible to show from the
  Boltzmann equation that the only possibility to have a vanishing
  collision term is to have $\mu=m$ and $\p_*=0$.}.  To account for
this possibility, let us replace the distribution $f(\p)$ by
\begin{equation}
f(\p)+n_c\,(2\pi)^3\delta(\p)\; ,
\label{eq:fsplit}
\end{equation}
with the understanding that the integral of $f(\p)$ in a small sphere
of radius $\epsilon$ around $\p=0$ goes to zero as $\epsilon\to
0$. This separation of the distribution into a continuous piece and a
singular piece at the origin is very useful when solving numerically
the Boltzmann equation, because of the impossibility to properly
represent a delta function after discretizing momentum space.

After this separation, one gets coupled equations for the evolution of
$f$ and $n_c$~\cite{LacazLPR1}. Formally, these can be derived by
injecting eq.~(\ref{eq:fsplit}) into the usual Boltzmann equation, and
by using the delta function that comes with $n_c$ in order to simplify
the integrals. It is important to note that terms that are at most
linear in $n_c$ can emerge from the collision term; the terms in
$n_c^2$ and $n_c^3$ all vanish because of kinematics or because the
gain and loss terms cancel. If we start from the complete $2\to 2$
collision term, we get the following two coupled equations
\begin{eqnarray}
\partial_t f(\p)
&&=
\smash{\frac{g^4}{4\omega_\p}\int_{\p'\k\k'}}
(2\pi)^4\delta(P+K-P'-K')\nonumber\\
&&
\qquad\qquad\qquad\qquad\times[f(\p')f(\k')(1+f(\p))(1+f(\k))\nonumber\\
&&
\qquad\qquad\qquad\qquad\quad-f(\p)f(\k)(1+f(\p'))(1+f(\k'))]
\nonumber\\
&&
+
\smash{\frac{g^4\,n_c}{8m\omega_\p}\int_{\p'\k'}}
(2\pi)^4\delta(\omega_\p+m-\omega_{\p'}-\omega_{\k'})\delta(\p-\p'-\k')\nonumber\\
&&
\qquad\qquad\qquad\qquad\times[f(\p')f(\k')(1+f(\p))\nonumber\\
&&
\qquad\qquad\qquad\qquad\quad-f(\p)(1+f(\p'))(1+f(\k'))]
\nonumber\\
&&
+
\smash{\frac{g^4\,n_c}{4m\omega_\p}\int_{\k\k'}}
(2\pi)^4\delta(\omega_\p+\omega_\k-m-\omega_{\k'})\delta(\p+\k-\k')\nonumber\\
&&
\qquad\qquad\qquad\qquad\times[f(\k')(1+f(\p))(1+f(\k))\nonumber\\
&&
\qquad\qquad\qquad\qquad\quad-f(\p)f(\k)(1+f(\k'))]\; ,
\end{eqnarray}
and
\begin{eqnarray}
\partial_t n_c
&&=
\smash{\frac{g^4\,n_c}{4m}\int_{\p'\k\k'}}
(2\pi)^4\delta(m+\omega_\k-\omega_{\p'}-\omega_{\k'})\delta(\k-\p'-\k')\nonumber\\
&&
\qquad\qquad\qquad\qquad\times[f(\p')f(\k')(1+f(\k))\nonumber\\
&&
\qquad\qquad\qquad\qquad\quad-f(\k)(1+f(\p'))(1+f(\k'))]\; .
\label{eq:coupled1}
\end{eqnarray}
It is straightforward to check that these coupled equations lead to
the following conservation laws~:
\begin{eqnarray}
n_c+\int\frac{d^3\p}{(2\pi)^3}\;f(\p)&=&{\rm const}\; ,
\nonumber\\
m\,n_c+\int\frac{d^3\p}{(2\pi)^3}\;\omega_\p\,f(\p)&=&{\rm const}\; ,
\nonumber\\
\int\frac{d^3\p}{(2\pi)^3}\;\p\,f(\p)&=&{\rm const}.
\label{eq:coupled2}
\end{eqnarray}

\section{Ultraviolet divergences in eq.~(\ref{eq:Cclass1})}
\label{sec:UVcoll}
In this section, we extract the sources of explicit dependence on the
ultraviolet cutoff in the classical approximations of the collision
term. By explicit, we mean the UV cutoff dependence even if one
assumes that the distribution $f(p)$ has a compact support and
therefore does not extend up to the cutoff.  

Firstly, let us note that there is no dependence on the UV cutoff
coming from the terms that are cubic and quadratic in the distribution
$f$ when $f(p)$ drops faster than $1/p$ at large $p$. Therefore, the
unapproximated collision term is UV finite. Likewise, the classical
approximation given in eq.~(\ref{eq:Cclass0}), that retains only the
cubic terms, has an UV finite collision term. However, as we shall see
later, the fixed point of the corresponding Boltzmann equation
acquires a dependence on the UV cutoff due to the fact that $f(p)$
converges towards a distribution that falls like $1/p$.

In the rest of this section, we will consider only the case of the
second form of the classical approximation, in which the collision
term is given by eq.~(\ref{eq:Cclass1}).

\subsection{Kubo-Martin-Schwinger symmetry}
Let us first stress a subtlety in the derivation of the formula
(\ref{eq:Cclass1}) from eq.~(\ref{eq:coll12}). As one can see,
eq.~(\ref{eq:Cclass1}) vanishes if the particle distribution is
identically zero. Physically, this is expected since this property
means that the collision rate is zero when there are no particles in the
system. Going back to eq.~(\ref{eq:coll12}), this property is
equivalent to the identity
\begin{equation}
\left[
{\bs\Sigma}_{11}(P)
+\frac{1}{2}({\bs\Sigma}_{21}(P)-{\bs\Sigma}_{12}(P))
\right]_{\rm vacuum} = 0\; .
\label{eq:KMSvac}
\end{equation}
This identity should be true because it is a consequence of the
Kubo-Martin-Schwinger \cite{Kubo1,MartiS1} (KMS) symmetry\footnote{In
  the Schwinger-Keldysh formalism, the KMS symmetry states that
  $\Sigma_{+-}(P)=\exp(-p_0/T)\Sigma_{-+}(P)$ for a system in
  equilibrium at the temperature $T$. Taking the limit $T\to 0^+$ and
  assuming that $p_0>0$, this leads to $\Sigma_{+-}^{\rm
    vacuum}(p_0>0)=0$, which is equivalent to eq.~(\ref{eq:KMSvac}).}.

It turns out that this identity is true in the classical statistical
approximation ${\cal C}^1$ as well, but in a rather subtle
way. Individually, ${\bs\Sigma}_{11}$ and
${\bs\Sigma}_{21}-{\bs\Sigma}_{12}$ are given by ultraviolet divergent
integrals, and make sense only when regularized with some cutoff
$\Lambda_{_{\rm UV}}$. When $P^2=0$ and $p_0>0$, we have (see the
appendix \ref{app:sigma} for some intermediate steps of this
calculation)
\begin{equation}
{\rm Im}\,{\bs\Sigma}_{11}(P)=\frac{I_++I-}{6}\quad,\quad
{\rm Im}\,{\bs\Sigma}_{21}(P)=\frac{I_+-I-}{2}
\label{eq:2loopS0}
\end{equation}
where
\begin{equation}
I_+=-\frac{g^4}{512\pi^3}\left(\Lambda_{_{\rm UV}}^2-\frac{2}{3}p^2\right)
\quad,\quad
I_-=-\frac{g^4}{256\pi^3}\left(\Lambda_{_{\rm UV}}^2-\frac{2}{3}p^2\right)\; .
\label{eq:2loopS}
\end{equation}

When combined into eq.~(\ref{eq:KMSvac}), these quantities cancel and
produce the expected identity. There is however a hidden complication:
this cancellation works only {\sl provided that one does not break the
  symmetry between the three internal lines of the 2-loop
  self-energy}. In deriving the eqs.~(\ref{eq:2loopS}), we have been
careful to use a regularization that does not to introduce any
asymmetry between the internal lines, but more naive regularizations
will in general break this identity. This is something to keep in mind
in numerical implementations of this approximation\footnote{The main
  issue is with ${\bs\Sigma}_{12}$, that contains one retarded
  $G_{_R}^0$ and two symmetric propagators $G_{_S}^0$. In some
  numerical implementations of the classical statistical
  approximation, the spectrum of the initial fluctuations is cut at a
  scale $\Lambda_{_{\rm UV}}$ which is smaller than the lattice
  ultraviolet cutoff. This amounts to limiting to $\Lambda_{_{\rm
      UV}}$ the momentum flowing through the symmetric propagators,
  while the momentum flowing through the retarded propagators can go
  all the way up to the lattice cutoff. In these implementations,
  eq.~(\ref{eq:KMSvac}) is not satisfied, and there is a non-zero
  collision rate even for an empty system.}.

\subsection{Terms quadratic in the UV cutoff}
Assuming that such a symmetric regularization has been employed, there
is no $f$-independent term in the classical collision term. Since the
terms that are cubic and quadratic in $f$ are UV finite, the
potentially UV divergent part of the collision term comes from the
terms that are linear in $f$,
\begin{eqnarray}
&&C_{\p,{\rm lin}}^{{\cal C}^1}[f]=\frac{g^4}{16\omega_\p}
\int_{\k}\int_{\p'}\int_{\k'}
(2\pi)^4\delta(P+K-P'-K')\;
\nonumber\\
&&\qquad\qquad\qquad\qquad\qquad\times
\big[f(\p')+f(\k')-f(\p)-f(\k)
\big]\; .
\label{eq:Cclass1-lin}
\end{eqnarray}
From this formula, an explicit calculation shows that the Boltzmann
equation in the classical approximation ${\cal C}^1$ contains a term
which is quadratic in the UV cutoff (see the appendix \ref{app:lin}
for the full expression of $C_{\p,{\rm lin}}^{{\cal C}^1}[f]$),
\begin{equation}
(\partial_t+\v_\p\cdot{\bs\nabla}) f(\p)
=-\frac{g^4\Lambda_{_{\rm UV}}^2}{1024\pi^3}\;\frac{1}{\omega_\p}\; f(\p) +\cdots
\label{eq:collUV}
\end{equation}
where the dots are terms that have a lesser degree in the UV cutoff
(including terms that are UV finite). Note that this term is a direct
consequence of the term quadratic in $\Lambda_{_{\rm UV}}$ contained
in ${\bs\Sigma}_{21}-{\bs\Sigma}_{12}$, that was exhibited in
ref.~\cite{EpelbGW1} and is a direct consequence of the
non-renormalizability of the classical statistical approximation.

\begin{figure}[htbp]
\begin{center}
\resizebox*{8cm}{!}{\includegraphics{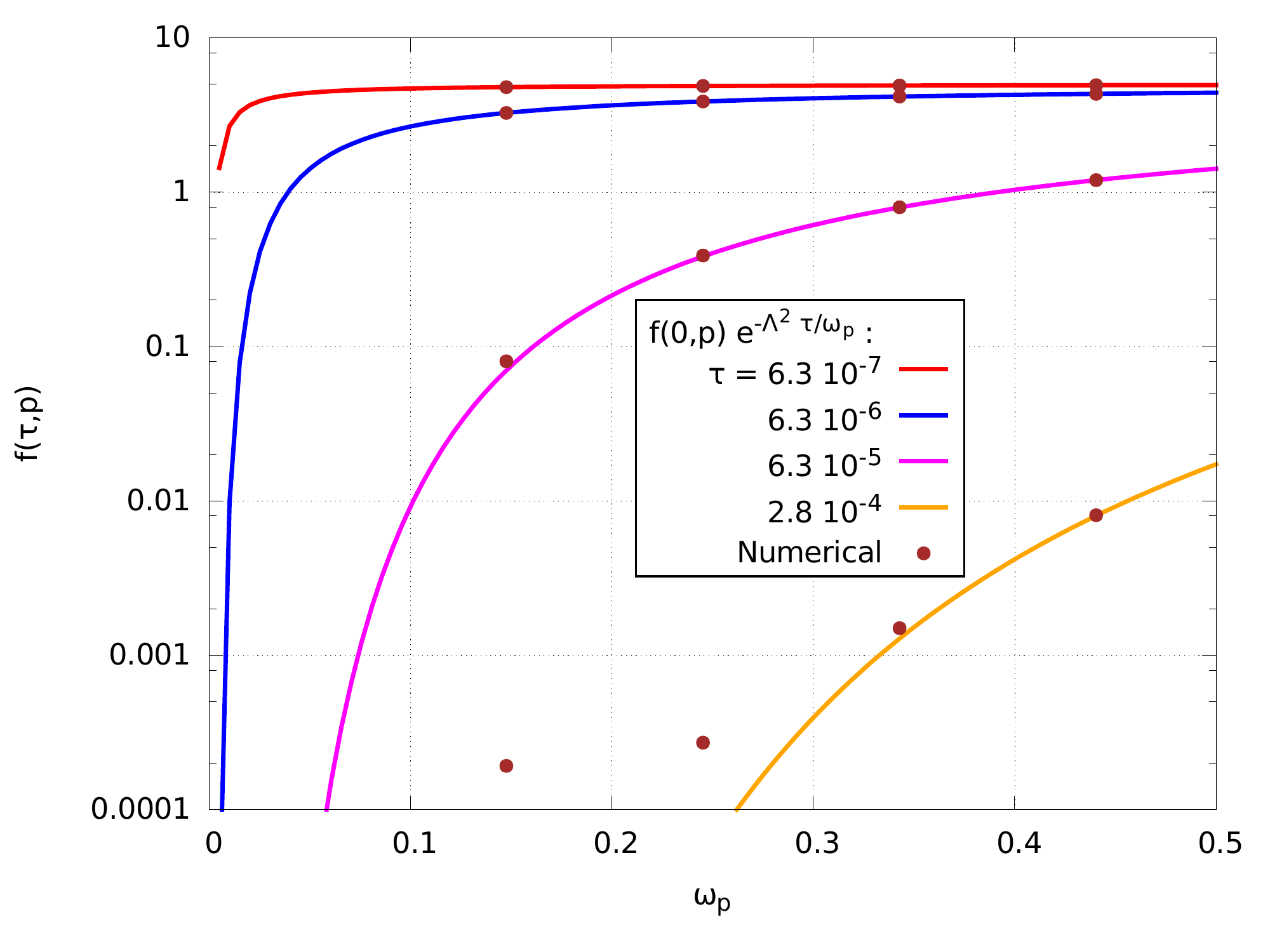}}
\end{center}
\caption{\label{fig:f0} Evolution of $f(\p)$ at short times in the
  classical approximation ${\cal C}^1$ for a large UV cutoff. Points~:
  numerical resolution of eq.~(\ref{eq:Cclass1}). Solid lines~: right
  hand side of eq.~(\ref{eq:fapprox}) (we denote $\tau\equiv g^4 t/(1024\pi^3)$). {Here, $\Lambda_{_{\rm UV}}$=100, $f_0 = 5.0$, $Q=0.5$ and $m=0.05$.}}
\end{figure}

This term in $\Lambda_{_{\rm UV}}^2$ introduces in the solution of
this Boltzmann equation a contamination by cutoff effects, no matter
what the initial $f(\p)$ is. In other words, even with an initial
$f(\p)$ that has a compact support and therefore does not touch the UV
cutoff, this term makes the solution immediately cutoff dependent (the
magnitude of the contamination, and how quickly it affects the
solution, of course depend on the coupling constant).  If one keeps
only this term and one assumes that the system is spatially
homogeneous, the solution would read~:
\begin{equation}
f(t,\p)
\approx
f(0,\p)
\;\exp\left[-\frac{g^4\Lambda_{_{\rm UV}}^2}{1024\pi^3}\;\frac{t}{\omega_\p}\right]\; .
\label{eq:fapprox}
\end{equation}
This formula indicates that the dependence on the UV cutoff leads to a
rapid depletion of the distribution at small momentum. This can be
tested by comparing this analytical formula with the numerical
solution of the Boltzmann equation with the collision term
(\ref{eq:Cclass1}), shown in the figure \ref{fig:f0} at four
different times shortly after the start of the evolution.
In this plot, one sees a very good agreement at short times between
the analytical estimate and the numerical solution of the Boltzmann
equation.  It shows that, when the UV cutoff is large, its effects
start at low momentum and grow exponentially in time. Note that the
estimate of eq.~(\ref{eq:fapprox}) is accurate only at very small
times (one can see deviations at the fourth of the times considered in
the figure \ref{fig:f0}), because of the neglected subleading terms in
the UV cutoff, and because of the terms in $f^3$ and $f^2$. At late
times, when the solution approaches a fixed point, all the terms in
the collision term are equally important and contribute to controlling
the fixed point.

From the analytical solution given in eq.~(\ref{eq:fapprox}), one can
estimate the timescale $t_*$ at which these spurious effects become
corrections of order one,
\begin{equation}
t_*^{-1}\sim \frac{g^4\Lambda_{_{\rm UV}}^2}{1024\pi^3 Q}\; ,
\end{equation}
where $Q$ is the typical physical momentum scale. Equivalently, this
can be written as
\begin{equation}
Qt_*\sim \frac{1024\pi^3}{g^4}\;\left(\frac{Q}{\Lambda_{_{\rm UV}}}\right)^2\; .
\label{eq:tstar}
\end{equation}
This timescale is reduced at larger couplings, or if the ultraviolet
cutoff is taken far above the physical scale. Note however that the
large numerical prefactor $1024\pi^3$ helps to mitigate these effects,
which suggests that $g=1$ is not a very strong coupling in this
theory.  In the following sections, we study the fixed points of the
Boltzmann equation and the behavior of the solutions at large times,
which means that we are well after this characteristic time. Our goal
is to assess how strong is the dependence on the UV cutoff when $t\gtrsim t_*$.

\section{H-theorem and asymptotic state}
\label{sec:asympt}
In the previous section, we have studied the explicit dependence on
the UV cutoff of the collision term in the classical
approximation. For distribution $f(\p)$ with a compact support, we
have seen that in the approximation ${\cal C}^0$ the collision term
does not depend on $\Lambda_{_{\rm UV}}$. This implies that at early
times, the evolution of $f(\p)$ is not affected by the cutoff
(provided that the cutoff is large enough for the support of $f(\p)$
to be below the cutoff). In contrast, in the approximation ${\cal
  C}^1$, the collision term contains a term which is quadratic in the
cutoff. This term leads to a sensitivity of $f(\p)$ on the cutoff even
at early times, which is more prominent if the cutoff is larger and at
small momentum $p$.

In this section, we now investigate the limit where time goes to
infinity. In the Boltzmann equation, the allowed asymptotic states and
the convergence of $f(\p)$ to a solution of that form can be proven on
the basis of conservation laws alone, and does not require any
numerical study. By doing so for the two classical approximations
considered in this paper, we will gain further insight on the effects
of the classical approximation.

\subsection{H-theorem}
\subsubsection{Collisional invariants}
The first thing to note about eqs.~(\ref{eq:Cclass0}) and
(\ref{eq:Cclass1}) is that the approximations performed on the
combinations of $f$'s that appear in the collision term do not affect
conservation laws, of particle number, energy and momentum.  Moreover,
the integrand in these approximate collision terms is still symmetric
under the following exchanges,
\begin{eqnarray*}
&& P\leftrightarrow K\\
&& P'\leftrightarrow K'\\
\end{eqnarray*}
and antisymmetric under the exchange
\begin{eqnarray*}
&& (P,K)\leftrightarrow (P',K')\; .
\end{eqnarray*}

A {\sl collisional invariant} $I(\p)$ is a quantity (possibly momentum
dependent) assigned to a particle, additive for an ensemble of
particles, that is conserved in elementary collisions~:
\begin{equation}
I(\p)+I(\k)=I(\p')+I(\k')\; .
\end{equation}
Thanks to the above symmetries, it is easy to check that the total
amount of this quantity in the system,
\begin{equation}
{\cal I}[f]\equiv\int\frac{\rmd^3\p}{(2\pi)^3}\;f(\p)\,I(\p)\; ,
\end{equation}
is constant in time if the particle distribution obeys the Boltzmann
equation, regardless of whether we use the non-approximated collision
term or the approximations of eqs.~(\ref{eq:Cclass0}) or
(\ref{eq:Cclass1})

\subsubsection{Full quantum case}
For the non-approximated collision term, it is well
known that one can define an entropy density,
\begin{equation}
S_{\rm quantum}=
\int\frac{\rmd^3\p}{(2\pi)^3}\;\left[(1+f(\p))\log(1+f(\p))-f(\p)\log(f(\p))\right]
\label{eq:Sfull}
\end{equation}
such that $\partial_t{S}_{\rm quantum}\ge 0$ for any (positive
definite) distribution\footnote{If the system is not spatially
  homogeneous, one must also introduce an entropy current ${\bs J}_s$,
  and the quantity which is always positive or zero is
  $\partial_t{S}_{\rm quantum}+{\bs\nabla}\cdot{\bs J}_s$.}
$f(\p)$. Furthermore, $\partial_t{S}_{\rm quantum}$ vanishes
(i.e. entropy stops being produced) provided that
$\log(f(\p)/(1+f(\p)))$ is a linear combination of the collisional
invariants allowed by the collision processes under consideration. For
the elastic collisions considered here, these are the four components
of the momentum, and the number of particles. This leads to the fact
that the most general fixed point of the Boltzmann equation is a
Bose-Einstein distribution with a chemical potential\footnote{ If one
  also includes number changing processes in the collision term, the
  reasoning remains unchanged until the last step: the number of
  particles is no longer conserved and thus cannot appear in the
  equilibrium value of $\log(f(\p)/(1+f(\p)))$. The allowed fixed
  points in this case are thus Bose-Einstein distributions with
  $\mu=0$.}.

\subsubsection{Classical approximation ${\cal C}^0$}
This reasoning can be generalized to the Boltzmann equation in which
the collision term has been approximated by eqs.~(\ref{eq:Cclass0}) or
(\ref{eq:Cclass1}), provided we modify the definition of the entropy
density. For the collision term (\ref{eq:Cclass0}), the
appropriate\footnote{What is ``appropriate'' is judged by whether the
  definition agrees with eq.~(\ref{eq:Sfull}) at large $f$ (to LO in
  the case of eq.~(\ref{eq:Cclass0}), and to LO and NLO in the case of
  eq.~(\ref{eq:Cclass1})), and whether it leads to a monotonously
  increasing entropy with time.} definition of the entropy density is
\begin{equation}
S({\cal C}^0)=
\int\frac{\rmd^3\p}{(2\pi)^3}\;\log(f(\p))\; .
\end{equation}

Using the Boltzmann equation, one finds that
\begin{eqnarray}
&&
\partial_t{S}({\cal C}^0)=\frac{g^4}{2}
\int_{\p}\int_{\k}\int_{\p'}\int_{\k'}
(2\pi)^4\delta(P+K-P'-K')\;
\nonumber\\
&&
\qquad\qquad\qquad
\times\frac{1}{\alpha_\p}
\big[
\alpha_{\p'}\alpha_{\k'}(\alpha_\p+\alpha_\k)
-
\alpha_{\p}\alpha_{\k}(\alpha_{\p'}+\alpha_{\k'})
\big]\; ,
\label{eq:H1}
\end{eqnarray}
where we denote $\alpha_\k\equiv f(\k)$. Using the
symmetries listed at the beginning of this subsection, this can be
rewritten as\begin{eqnarray}
&&
\partial_t{S}({\cal C}^0)=\frac{g^4}{8}
\int_{\p}\int_{\k}\int_{\p'}\int_{\k'}
(2\pi)^4\delta(P+K-P'-K')\;
\nonumber\\
&&
\qquad
\times
\Big[
\frac{1}{\alpha_\p}
+
\frac{1}{\alpha_\k}
-
\frac{1}{\alpha_{\p'}}
-
\frac{1}{\alpha_{\k'}}
\Big]
\big[
\alpha_{\p'}\alpha_{\k'}(\alpha_\p+\alpha_\k)
-
\alpha_{\p}\alpha_{\k}(\alpha_{\p'}+\alpha_{\k'})
\big]\; ,
\label{eq:H2}
\end{eqnarray}
and it is now easy to check that the integrand is positive definite,
which generalizes the H-theorem to the classical statistical
approximation ${\cal C}^0$ of the Boltzmann equation. Thanks to the
factor
\begin{equation}
\frac{1}{\alpha_\p}
+
\frac{1}{\alpha_\k}
-
\frac{1}{\alpha_{\p'}}
-
\frac{1}{\alpha_{\k'}}
\end{equation}
in the integrand, we also conclude that the fixed point is reached if
$1/\alpha_\p$ is a linear combination of collisional invariants, which
for elastic collisions gives the following most general equilibrium
distribution,
\begin{equation}
f_{{\cal C}^0}(\p)=\frac{T}{\omega_\p-\mu}\; .
\label{eq:class-eq0}
\end{equation}
Note that this expression is made of the first term in the expansion
of the Bose-Einstein distribution in the limit of small $\omega_\p$.
Unlike the full Bose-Einstein distribution that has an exponentially
falling tail, this equilibrium distribution decreases only as a power
law. As we will see in the next section, when inserted into the
equations of particle number and energy conservation, this leads to
parameters $T$ and $\mu$ that depend on the UV cutoff.

\subsubsection{Classical approximation ${\cal C}^1$}
In the case of the collision term (\ref{eq:Cclass1}), one should use
\begin{equation}
S({\cal C}^1)=
\int\frac{\rmd^3\p}{(2\pi)^3}\;\log\left(\frac{1}{2}+f(\p)\right)\; .
\end{equation}
The reasoning is exactly the same as in the case of the approximation
${\cal C}^0$, and the equations (\ref{eq:H1}) and (\ref{eq:H2}) remain
valid for the derivative $\partial_t{S}({\cal C}^1)$ provided that we now
define $\alpha_\k\equiv f(\k)+\frac{1}{2}$. The fixed points are still
obtained by imposing that $1/\alpha_\p$ be a linear combination of
collisional invariants. This  gives~:
\begin{equation}
f_{{\cal C}^1}(\p)=\frac{T}{\omega_\p-\mu}-\frac{1}{2}\; ,
\label{eq:class-eq1}
\end{equation}
which now corresponds to the first {\sl two} orders in the expansion
of the Bose-Einstein distribution at small energy.

To conclude this subsection, let us note that when one includes only the
$2\to 2$ scattering in the collision term, all the coupling dependence
factorizes in the  factor $g^4$ that appears in the right hand side of
the Boltzmann equation. One can therefore absorb this factor into a
rescaling of the time axis, which means that the only effect of
changing the coupling is to make the time evolution faster or slower
(timescales are proportional to $g^{-4}$).  But the outcome of the
evolution at $t\to \infty$ does not depend on the value of the
coupling, nor the above conclusion that the particle distribution will
be driven eventually towards the functional form of
eqs.~(\ref{eq:class-eq0}) or (\ref{eq:class-eq1}), depending of which
version of the classical approximation is used.

\subsection{Dependence  of the fixed point on the UV cutoff}
The fixed point of the Boltzmann equation has a number of free
parameters equal to the number of conserved quantities in the system,
and therefore it is sufficient to write these conservation equations
in order to determine these parameters. In the case of elastic
collisions, in the classical approximation ${\cal C}^1$, $T$ and $\mu$
are related to the particle density and the energy density,
\begin{eqnarray}
n&=&n_c+\int\frac{\rmd^3\p}{(2\pi)^3}\; \left(\frac{T}{\omega_\p-\mu}-\frac{1}{2}\right)\nonumber\\
\epsilon&=&n_c m+\int\frac{\rmd^3\p}{(2\pi)^3}\; \left(\frac{T}{\omega_\p-\mu}-\frac{1}{2}\right)\;\omega_\p\; .
\label{eq:T-mu}
\end{eqnarray}
Note that the chemical potential must be less than (or equal to) the
mass of the particles for these integrals to be defined in the
infrared. In these equations, the constant $n_c$ accounts for the
possibility to form a Bose-Einstein condensate if the system is
overoccupied.

The main issue with these equations is that the integrals in both of
them are severely ultraviolet divergent. Let us therefore introduce an
ultraviolet cutoff such that $|\p|\le \Lambda_{_{UV}}$; the integrals
are now finite but depend also on $\Lambda_{_{UV}}$, which implies
that $T$, $\mu$ and $n_c$ will not only depend on the input quantities
$n$ and $\epsilon$, but also on the cutoff. In fact, if
$\Lambda_{_{UV}}$ is much larger than the physical scales, then the
values of $T$ and $\mu$ are completely dominated by the cutoff. This
property of the Boltzmann equation in the classical approximation is
closely related to the fact that the underlying field theory in this
approximation is not renormalizable~\cite{EpelbGW1}.

\begin{figure}[htbp]
\begin{center}
\resizebox*{8cm}{!}{\includegraphics{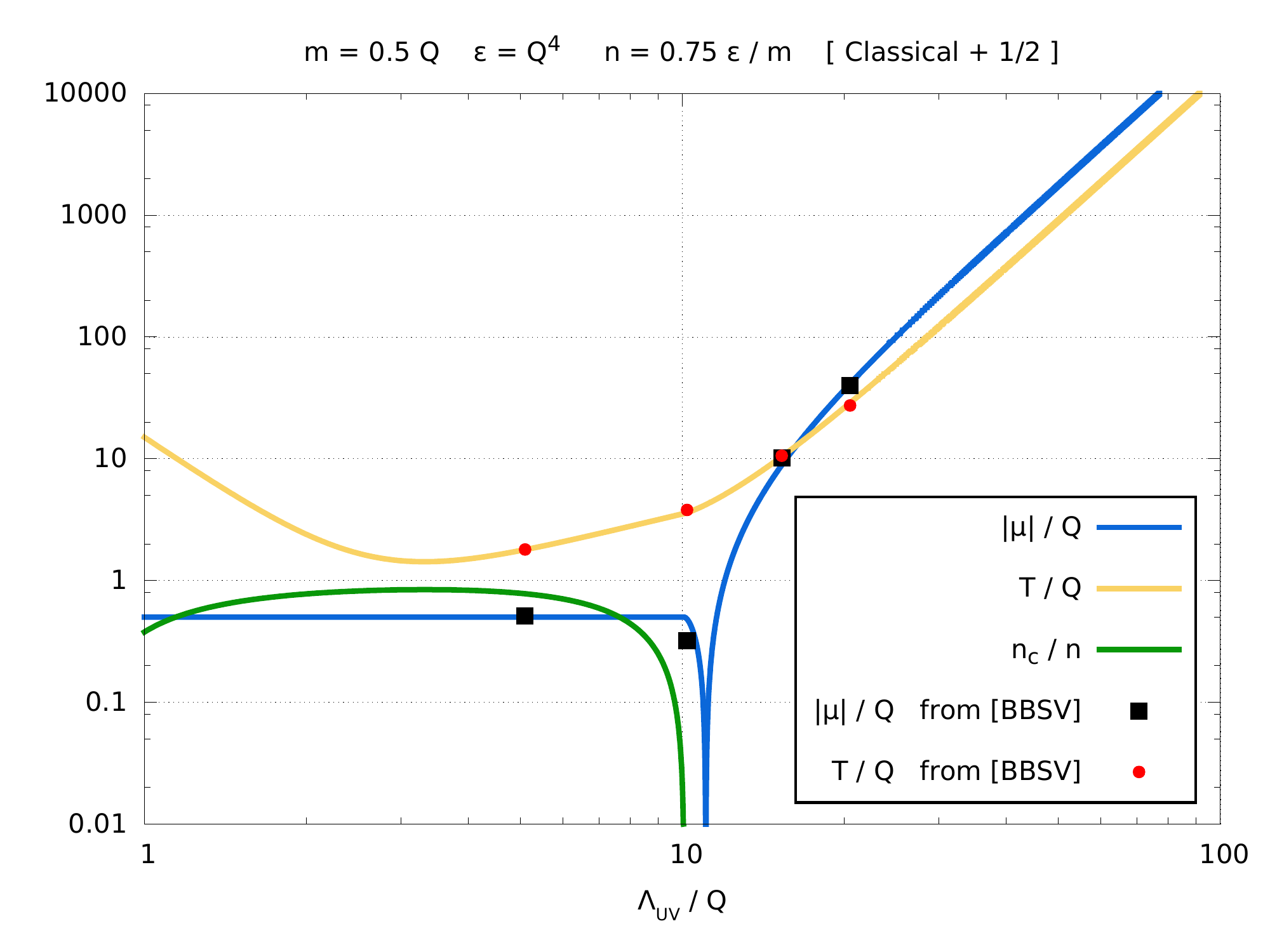}}
\end{center}
\caption{\label{fig:Tmu}Evolution of the classical equilibrium
  parameters $T,\mu,n_c$ as a function of the ultraviolet cutoff, for
  the classical approximation ${\cal C}^1$. All the dimensionful
  quantities in the system are set in terms of a unique parameter
  $Q=\epsilon^{1/4}$~: in this example, $m=0.5\cdot Q$ and
  $n=0.75\cdot \epsilon/m$. $\mu$ is positive at small
  $\Lambda_{_{UV}}$ and negative at large $\Lambda_{_{UV}}$. The
  points reproduce the values listed in the figure 10 of
  Ref.~\cite{BergeBSV3}, obtained with a classical statistical field
  simulation on a $768^3$ lattice.}
\end{figure}

Let us nevertheless use eqs.~(\ref{eq:T-mu}) with an ultraviolet
cutoff $\Lambda_{_{UV}}$. They can be rewritten as
\begin{eqnarray}
2\pi^2 ({\wt n}-{\wt n}_c) +\frac{1}{6}&=&
{\wt T}\int_0^1 \rmd{x}\;\frac{x^2}{\sqrt{x^2+{\wt m}^2}-\wt{\mu}}
\nonumber\\
2\pi^2 ({\wt\epsilon}-{\wt n}_c{\wt m}) +f({\wt m})&=&
{\wt T}
\int_0^1 \rmd{x}\;\frac{x^2\sqrt{x^2+{\wt m}^2}}{\sqrt{x^2+{\wt m}^2}-\wt{\mu}}\; ,
\label{eq:class-eq}
\end{eqnarray}
where we denote
\begin{equation}
f({\wt m})
\equiv
\frac{1}{8}\left(
\sqrt{1+{\wt m}^2}(1+\frac{{\wt m}^2}{2})
-\frac{{\wt m}^4}{2} \sinh^{-1}\frac{1}{{\wt m}}
\right)
\end{equation}
and where, for any quantity $X$ with mass dimension $d$, we denote
${\wt X}\equiv X/\Lambda_{_{UV}}^d$. These equations define the
asymptotic values\footnote{It may seem at first sights that there are
  three unknowns (${\wt T}, {\wt \mu}$ and ${\wt n}_c$) for only two
  equations. However, this is not the case. If the system is not
  overoccupied, then ${\wt n}_c=0$ and there are only two unknowns. If
  the system is overoccupied, then we know a priori that ${\wt
    \mu}={\wt m}$, and again only two of these parameters are truly
  unknown. {A discussion of the condition for the existence of a
    Bose-Einstein condensate is presented in the appendix
    \ref{app:masslessnc}.}} ${\wt T}, {\wt \mu}$ and ${\wt n}_c$ as a
function of ${\wt \epsilon}$, ${\wt n}$ and ${\wt m}$. If the cutoff
is large compared to the physical momentum scale, then we have
\begin{equation}
{\wt\epsilon}, {\wt n},{\wt m}\ll 1\; .
\end{equation}
Neglecting ${\wt m}\ll 1$ in eqs.~(\ref{eq:class-eq}), these equations
become
\begin{eqnarray}
2\pi^2 ({\wt n}-{\wt n}_c) +\frac{1}{6}&=&
{\wt T}\Big[
\frac{1}{2}+{\wt \mu}+{\wt\mu}^2\ln\left(\frac{1-{\wt\mu}}{-{\wt\mu}}\right)
\Big]
\nonumber\\
2\pi^2 {\wt\epsilon} +\frac{1}{8}&=&
{\wt T}
\Big[
\frac{1}{3}+\frac{{\wt\mu}}{2}+{\wt \mu}^2+{\wt\mu}^3\ln\left(\frac{1-{\wt\mu}}{-{\wt\mu}}\right)
\Big]\; ,
\label{eq:class-eq11}
\end{eqnarray}
and for ${\wt\epsilon}, {\wt n}\ll 1$ their solutions are
approximately given by
\begin{equation}
{\wt n}_c=0
\quad,\quad
{\wt\mu}\approx\frac{1}{80\pi^2(4{\wt\epsilon}-3{\wt n})}\to -\infty\quad,\quad
{\wt T}\approx -\frac{\wt\mu}{2}\to +\infty\; .
\end{equation}
Note that these values go to infinity very quickly if the ultraviolet
cutoff $\Lambda_{_{UV}}$ goes to infinity at fixed $\epsilon$ and $n$
(asymptotically, $\mu,T\sim \Lambda_{_{UV}}^4$). This behavior of the
asymptotic $T$ and $\mu$ is quite consistent with the observation made
in the figure 10 of Ref.~\cite{BergeBSV3}, although there the
classical statistical approximation for the underlying field theory
was used instead of the classical approximation of the Boltzmann
equation. At finite ultraviolet cutoff, the eqs.~(\ref{eq:class-eq})
can be solved numerically, as illustrated in the figure \ref{fig:Tmu}.
As one can see on this figure, the values of the equilibrium
parameters $T,\mu$ and $n_c$ depend on the ultraviolet cutoff,
especially strongly at large values of this cutoff.  Even the
conclusion regarding the existence of a Bose-Einstein condensate is
cutoff dependent, since the condensate density $n_c$ vanishes above a
certain value of the cutoff. When the cutoff is too small, some modes
that are important in the dynamics of the system are excluded from the
description. On the contrary, when the cutoff is large, the parameters
of the fixed point are controlled by the ultraviolet divergence in the
integrals of eqs.~(\ref{eq:T-mu}).  Moreover, because the fixed point
of the Boltzmann equation is independent of the value of the coupling,
the system will eventually evolve towards these unphysical values no
matter how small the coupling is. The only effect of a smaller
coupling is to make the convergence towards this asymptotic state
slower. Note also that there seem to be a ``sweet spot'', in the range
$3\lesssim \Lambda_{_{UV}}/Q\lesssim 6$, where these parameters are
the least sensitive to the value of the cutoff.

In the figure \ref{fig:Tmu}, we also show the comparison with the
values of $T,\mu$ fitted in Ref.~\cite{BergeBSV3} from several
classical statistical simulations with various ultraviolet cutoffs.
The remarkable agreement\footnote{It is important to keep in mind that
  the full fledged classical statistical simulations contain some
  effects of inelastic processes, in contrast to the Boltzmann
  equation studied here. Therefore, in the limit of extremely large
  times, these simulations should always lead to a vanishing chemical
  potential. However, on the timescales considered in
  Ref.~\cite{BergeBSV3}, this is not yet the case because they are
  still small compared to the inverse of the inelastic scattering
  rate.} with the asymptotic behavior of the elastic Boltzmann
equation in the classical approximation suggests that this much
simpler equation is a very good model in order to perform studies of
the range of validity of the classical statistical
approximation. There are two advantages in using the Boltzmann
equation for this purpose: (i) it is simpler to solve that the
classical statistical field simulation\footnote{As recalled in the
  appendix \ref{app:coll}, the collision term for elastic collisions
  in $\phi^4$ theory can be reduced to a 2-dimensional integral if the
  distribution is isotropic in momentum space.} and (ii) one can
compute the solutions with and without the classical approximation
equally easily. Given this, our goal in the rest of this paper is to
address the following questions:
\begin{itemize}
\item Compare the full Boltzmann equation and the Boltzmann equation
  in the classical statistical approximation, by solving them
  numerically. Since changes of the coupling can be absorbed into a
  rescaling of time, we will do this at a unique coupling $g^2=1$.
\item Assess how the value of the ultraviolet cutoff influences the
  results in the classical statistical approximation. In particular,
  is there a range of values of the cutoff that helps to reduce the
  differences between the full and approximated Boltzmann equations?
\end{itemize}

\begin{figure}[htbp]
\begin{center}
\resizebox*{8cm}{!}{\includegraphics{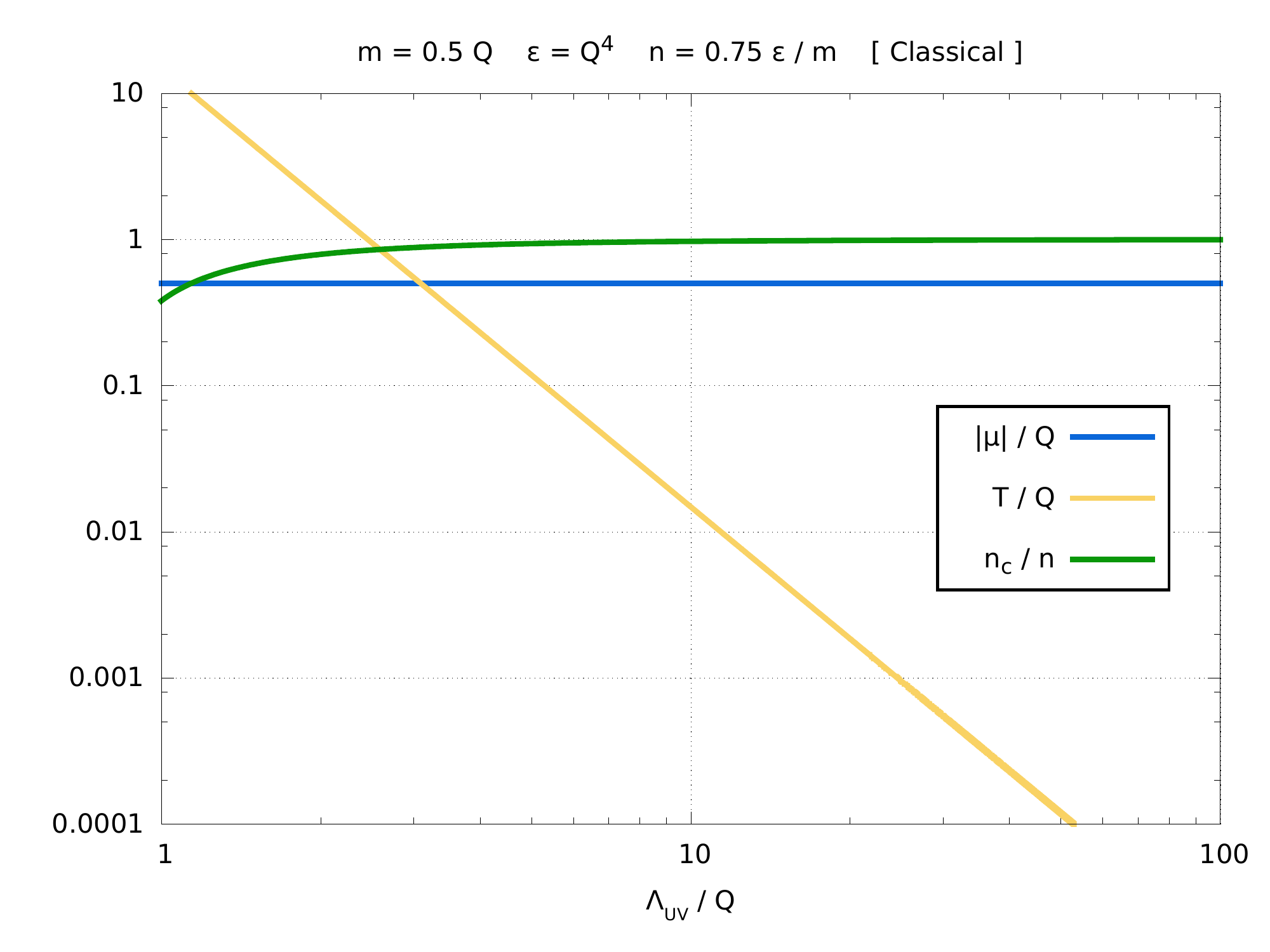}}
\end{center}
\caption{\label{fig:Tmu1}Evolution of the classical equilibrium
  parameters $T,\mu,n_c$ as a function of the ultraviolet cutoff, for
  the classical approximation ${\cal C}^0$. The setup is the same as
  in the figure \ref{fig:Tmu}.}
\end{figure}

Before going into this, let us also show how the parameters of the
fixed point depend on the UV cutoff in the case of the classical
approximation ${\cal C}^0$. In this case, the two conservation
equations lead to
\begin{eqnarray}
n&=&n_c+\int\frac{\rmd^3\p}{(2\pi)^3}\; \left(\frac{T}{\omega_\p-\mu}\right)\nonumber\\
\epsilon&=&n_c m+\int\frac{\rmd^3\p}{(2\pi)^3}\; \left(\frac{T}{\omega_\p-\mu}\right)\;\omega_\p\; ,
\label{eq:T-mu1}
\end{eqnarray}
which can easily be solved numerically. For the same set of input
parameters as in the figure \ref{fig:Tmu}, one now finds the results
of the figure \ref{fig:Tmu1}.
In this case, the temperature drops to zero as $\Lambda_{_{\rm
    UV}}^{-3}$ when the cutoff becomes large, while the chemical
potential always goes to $\mu=m$.  In this approximation, all the
particles go into a Bose-Einstein condensate in the limit
$\Lambda_{_{\rm UV}}\to \infty$.  This version of the classical
approximation is rather subtle because, as we have mentioned before,
the collision term is independent of $\Lambda_{_{\rm UV}}$ provided
that the distribution $f(\p)$ has a compact support that does not
extend to the cutoff. Therefore, it seems that for a sufficiently
large cutoff, the solution should not depend on the cutoff. However,
in this classical approximation, the limits $\Lambda_{_{\rm UV}}\to
\infty$ and $t\to\infty$ do not commute. The figure \ref{fig:Tmu1}
amounts to taking the limit $t\to\infty$ first, and to look at the
behavior of the resulting equilibrium parameters for a large
$\Lambda_{_{\rm UV}}$.

\section{Numerical study of the Boltzmann equation}
\label{sec:evol}
\subsection{Setup}
In order to go beyond the simple estimates and considerations based on
conservation laws that we have performed in the previous sections, we
now solve numerically the Boltzmann equation, in order to have access
to the details of the time evolution. In order to do so, we must do
some simplifying assumptions. Firstly, we assume that the system is
spatially homogeneous in order to drop the $\x$ dependence in the
distribution function. After this simplification, the collision term
is still given by a 5-dimensional integral (after having taken into
account energy and momentum conservation), which is still too
computationally expensive. Therefore, we further assume that the
distribution is isotropic in momentum space, which allows a drastic
simplification of the collision term: all the angular integrations can
be performed analytically, and the 5-dimensional integral is reduced
to a 2-dimensional integral, which is now doable numerically. This
reduction is well known\footnote{Note that this simplification is
  specific to point-like scalar interactions. For more general
  scatterings, the matrix element may have a non-trivial angular
  dependence, preventing this analytical simplification.}, but we
reproduce it in the appendix \ref{app:coll} for the sake of
completeness.

When solving numerically the Boltzmann equation, special care must be
taken to satisfy with high accuracy the conservation of particle
number, energy and momentum. This is particularly important in the
vicinity of a Bose-Einstein condensation transition, where the system
evolves very rapidly and where these errors may grow in an
uncontrollable manner. In the appendix \ref{app:coll}, we describe a
discretization scheme with which these conservation laws remain exact
(and thus are only limited by rounding errors in practice).

In this study, we are interested in initial distributions that are
large for momenta below a certain characteristic scale $Q$, and
negligible above $Q$. For simplicity, we take an initial distribution
of the form
\begin{equation}
f(0,p)=f_0\;\theta(Q-p)\; .
\label{eq:step}
\end{equation}
$Q$ is not a true parameter of the problem, since all the dimensionful
quantities can be expressed in units of $Q$. Likewise, the coupling
constant appears only as a $g^4$ prefactor in the collision term, and
its sole effect is to stretch or squeeze the timescales. We have
therefore taken $g=1$ in all the numerical calculations. Starting with
the same initial condition, we will compare the unapproximated
Boltzmann equation with the classical approximations ${\cal C}^0$ and
${\cal C}^1$, for several values of the ratio $\Lambda_{_{\rm UV}}/Q$.

If $f_0$ is large enough, the system becomes overoccupied, which leads
to the formation of a Bose-Einstein condensate. If one just solves the
plain Boltzmann equation (with or without the classical approximation
in the collision term), the solution becomes unstable at the time
where the BEC transition would normally happen.  In order to be able
to pursue the numerical evolution beyond this transition, it is
necessary to explicitly allow the presence of a condensate by solving
the coupled equations described in the section \ref{sec:bec}.  When
doing so, the region of the transition remains difficult to handle
numerically, because the density of condensed particles grows by
several orders of magnitude in a very short time. A very small
timestep is necessary in this region. In order to avoid complications
in the equations (\ref{eq:coupled1}) and (\ref{eq:coupled2}), we
consider massive particles (in the numerical results presented later
in this section, we have used $m=0.1 Q$). 

\begin{figure}[htbp]
\begin{center}
\resizebox*{8cm}{!}{\includegraphics{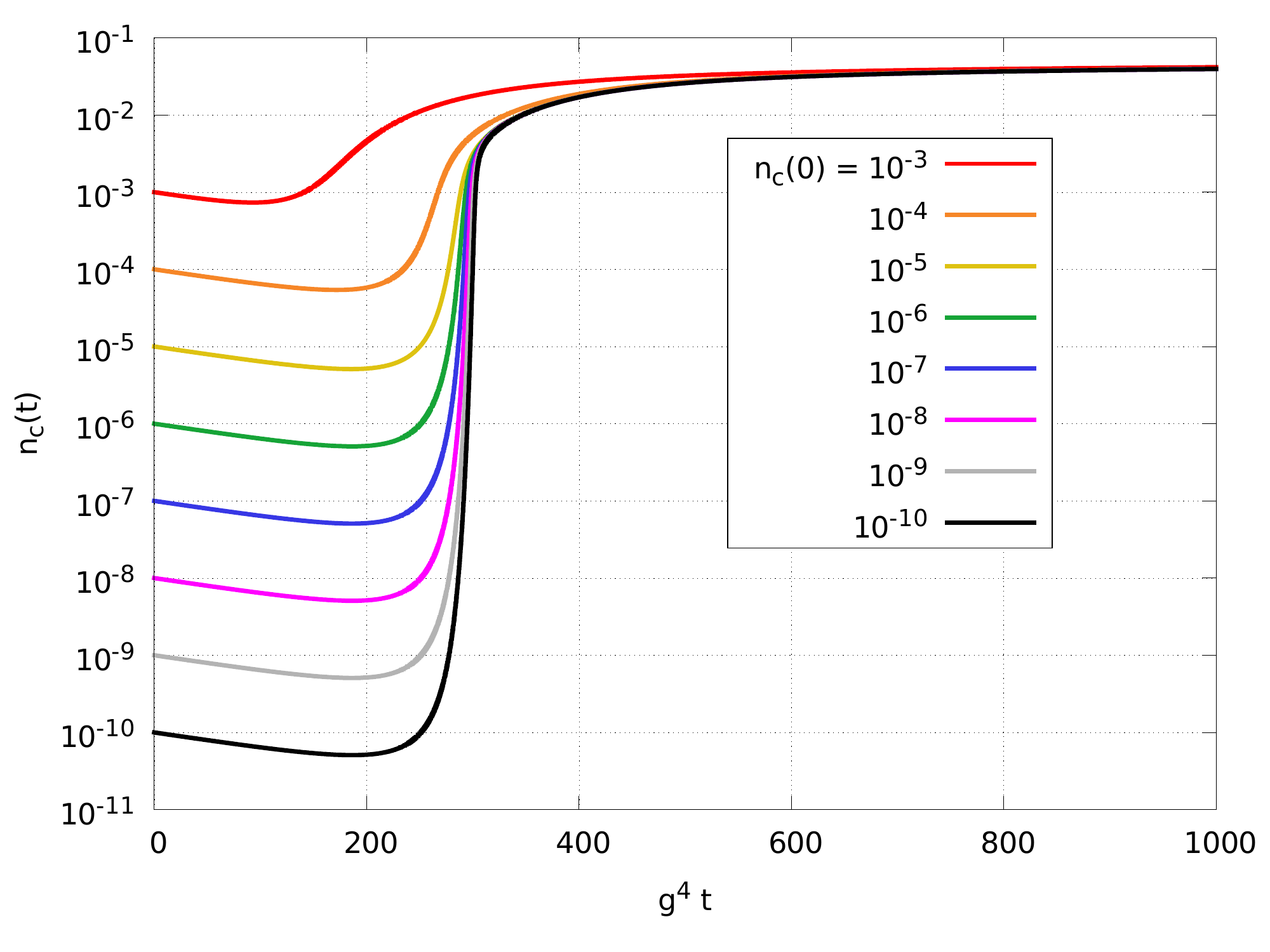}}
\end{center}
\caption{\label{fig:n0-dep}Time evolution of the density $n_c$ of
  condensed particles with the unapproximated collision term, for
  $n_c(0)$ varying from $10^{-3}$ to $10^{-10}$. The UV cutoff was set
  to $\Lambda_{_{\rm UV}}/Q=4$, and $f_0=5$.}
\end{figure}

From eq.~(\ref{eq:coupled2}), it is obvious that the initial value of
$n_c$ must be non-zero for $n_c$ to have an effect on the system,
since its evolution equation is homogeneous. However, the precise
value of this initial condition is not very important, as long as
$n_c(0)$ is negligible compared to the total particle density. This is
illustrated in the figure \ref{fig:n0-dep}, where we have varied the
initial value $n_c(0)$ by many orders of magnitude, at fixed
$f_0=5$. For a small enough $n_c(0)$, the time at which the BEC
transition occurs and the subsequent evolution of $n_c(t)$ do not
depend on this initial value.
In the rest of the paper, all the computations are done with the
initial value $n_c(0)=10^{-5}$.

\begin{figure}[htbp]
\begin{center}
\resizebox*{8cm}{!}{\includegraphics{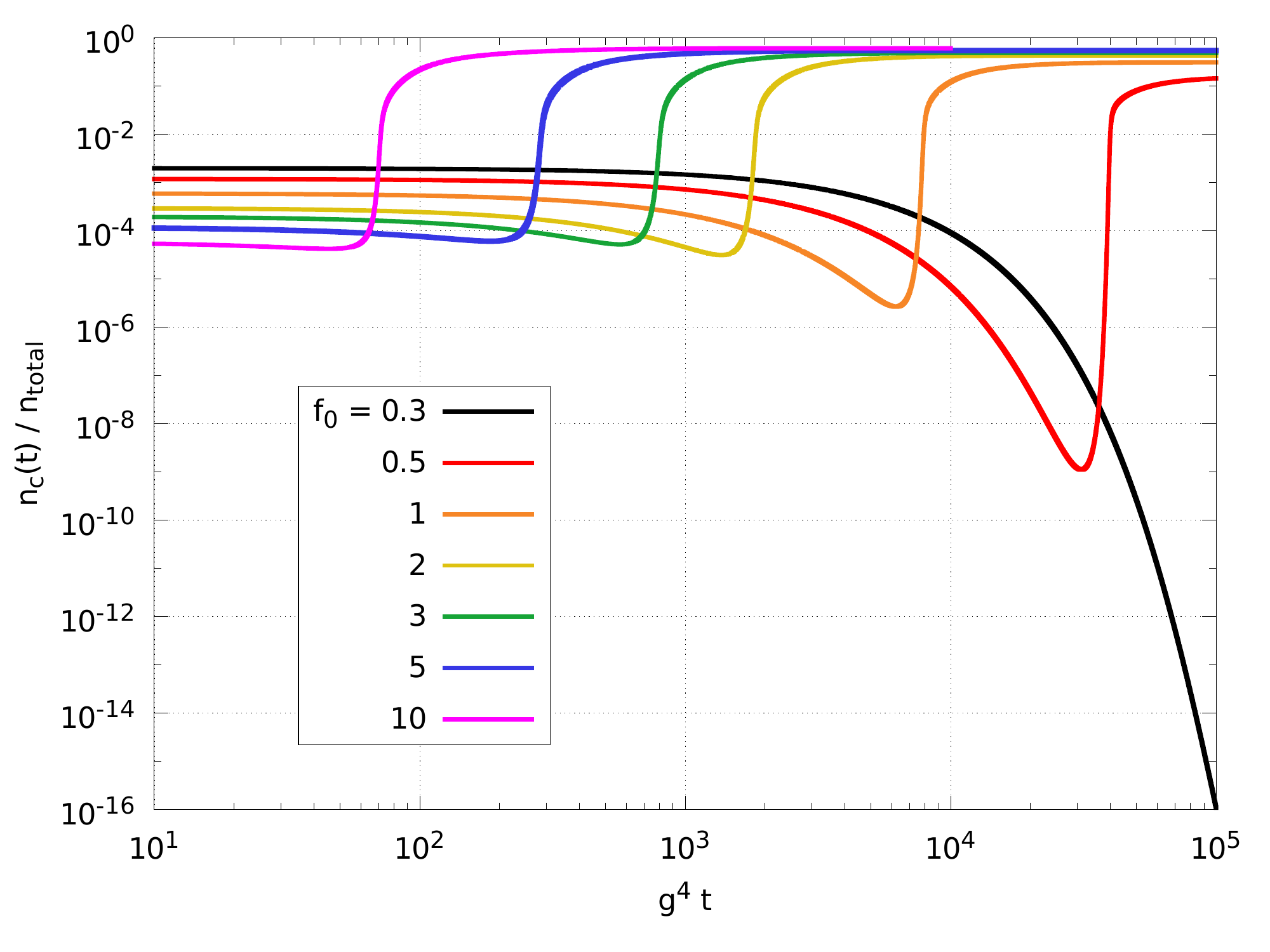}}
\end{center}
\caption{\label{fig:f0-dep}Time evolution of the fraction of particles
  in the zero mode, $n_c(t) / n_{\rm total}$, with the unapproximated
  collision term, for $f_0=0.3,0.5,1,2,3,5,10$. The UV cutoff was set
  to $\Lambda_{_{\rm UV}}/Q=4$.}
\end{figure}

In the figure \ref{fig:f0-dep}, we show how the time evolution of the
fraction of condensed particles $n_c(t)/n_{\rm total}$ changes with
the magnitude $f_0$ of the initial particle distribution.
The behavior of this ratio is quite sensitive to the value of $f_0$,
and a BEC transition happens for any $f_0$ larger than some critical
value that one can estimate to be $f_0^*\approx 0.30484$. Moreover, by
increasing $f_0$, the transition happens at earlier times, while the
asymptotic value of the condensed fraction increases only
slightly\footnote{Increasing $f_0$ is not an efficient way to increase
  the condensate fraction. Indeed, by spreading the extra particles
  uniformly at all momenta from $0$ to $Q$, one is mostly heating up
  the system.}. A more thorough description of the BEC transition in
kinetic theory can be found in refs.~\cite{SemikT1,SemikT2,LacazLPR1}.

\subsection{Evolution of the condensate density}
Firstly, let us consider as a reference the case where we do not
perform any classical approximation on the collision term. The initial
condition we have used is of the form given in eq.~(\ref{eq:step}),
with $f_0=10$.
\begin{figure}[htbp]
\begin{center}
\resizebox*{8cm}{!}{\includegraphics{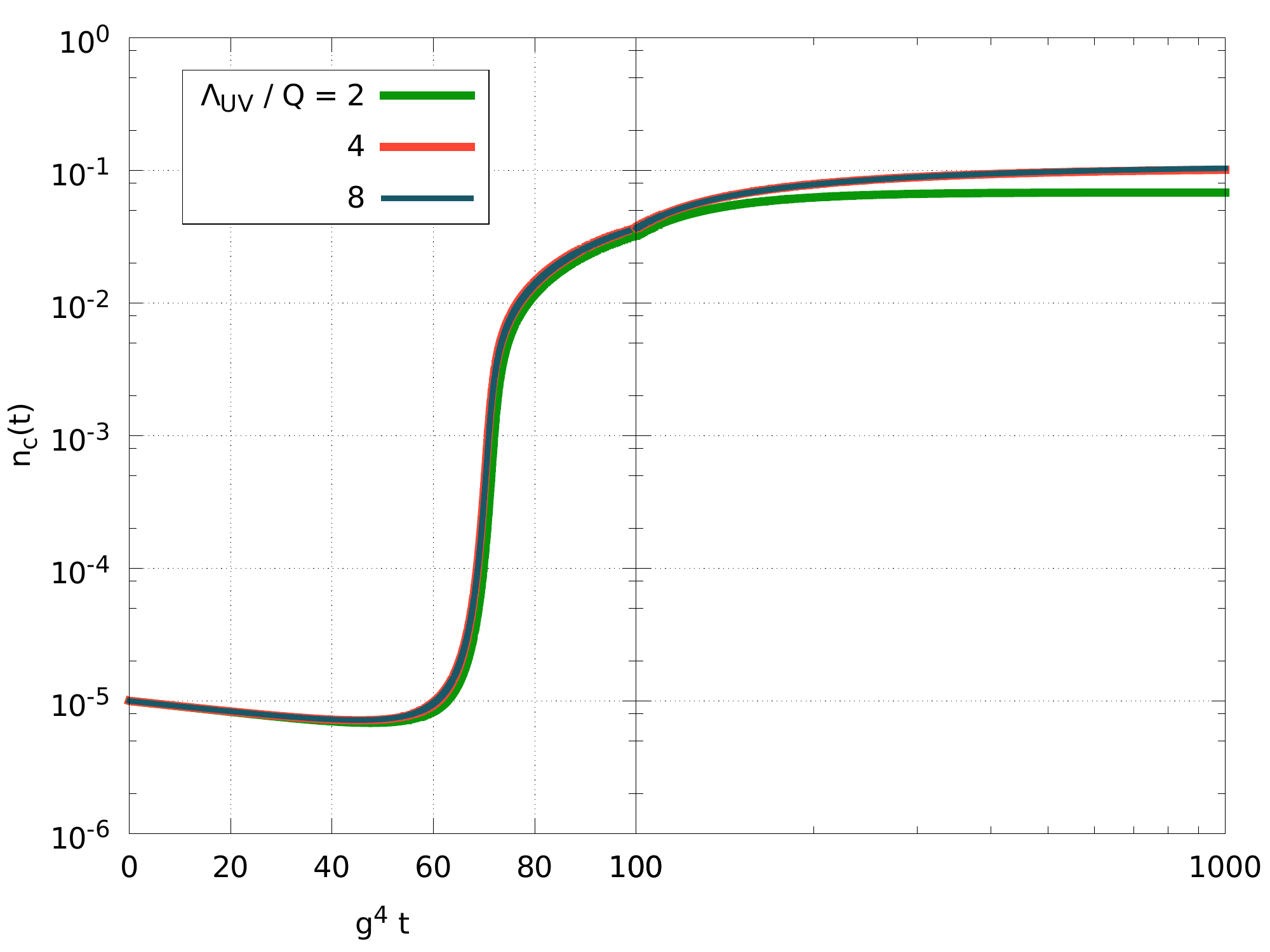}}
\end{center}
\caption{\label{fig:nc-Q}Time evolution of the density $n_c$ of
  condensed particles with the unapproximated collision term, for
  three values $\Lambda_{_{\rm UV}}/Q=2,4$ and $8$ of the UV cutoff.}
\end{figure}
In this case, the solution of the Boltzmann equation has a well
defined $\Lambda_{_{\rm UV}}\to\infty$ limit, and the only requirement
to observe this limiting behavior is to make the UV cutoff large
enough compared to the physical scale. In the figure \ref{fig:nc-Q},
we illustrate this by showing the time dependence of the density $n_c$
of condensed particles (for an overoccupied initial condition), for
 three values $\Lambda_{_{\rm UV}}/Q=2,4$ and $8$ of the UV
cutoff. We observe that the asymptotic value of $n_c$ changes a bit
when going from $\Lambda_{_{\rm UV}}/Q=2$ to $4$, but remains
unchanged when increasing it further to $8$. This is the sign that the
UV limit is already reached for values of the cutoff above $4Q$. In
the following, we will use this result as a reference, and refer to it
as the ``exact'' result.

\begin{figure}[htbp]
\begin{center}
\resizebox*{8cm}{!}{\includegraphics{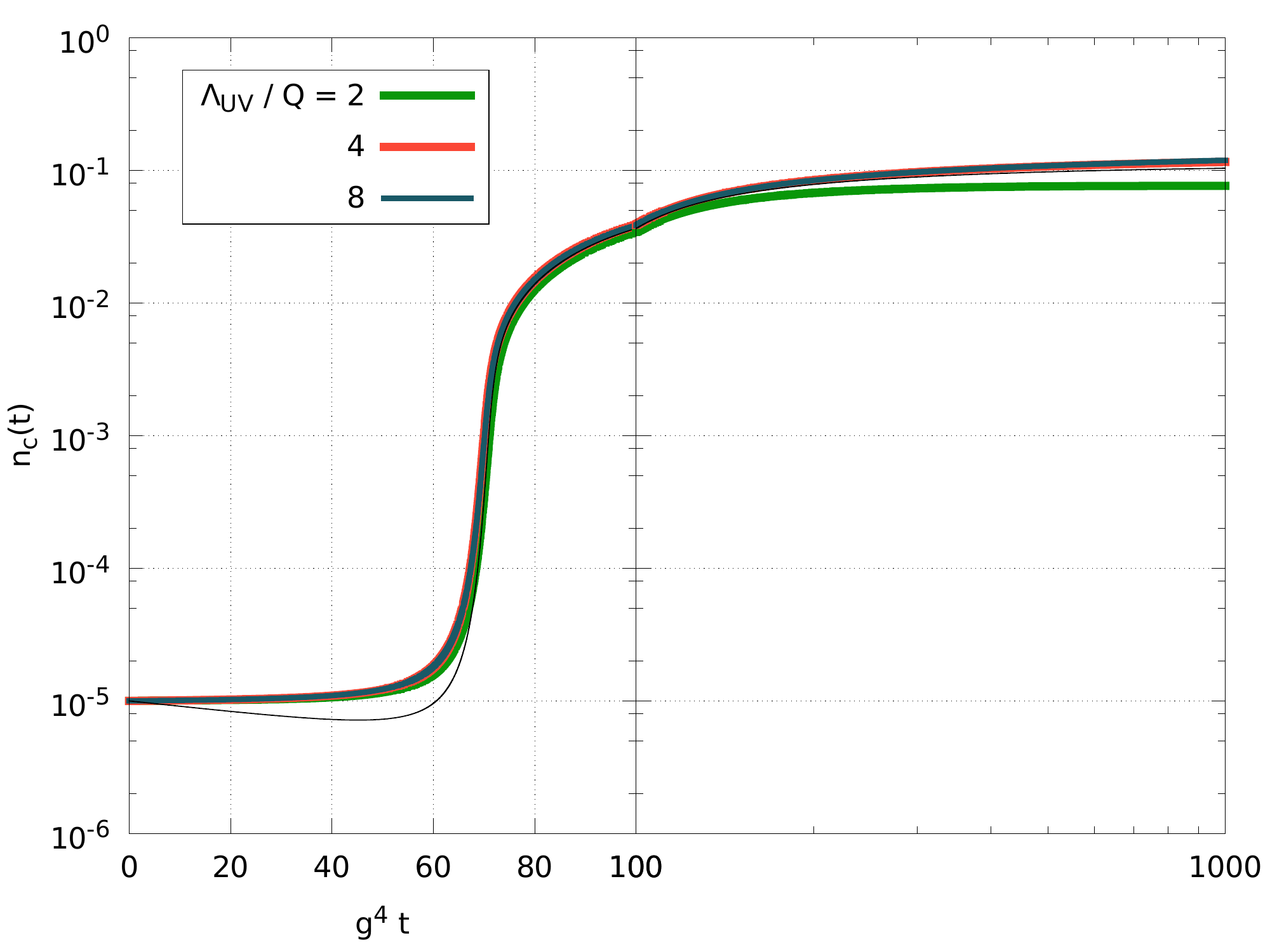}}
\end{center}
\caption{\label{fig:nc-C0}Time evolution of the density $n_c$ of
  condensed particles with the collision term in the classical
  approximation ${\cal C}^0$, for three values $\Lambda_{_{\rm
      UV}}/Q=2,4$ and $8$ of the UV cutoff. The thin black line is the
  solution of the unapproximated Boltzmann equation with
  $\Lambda_{_{\rm UV}}/Q=8$.}
\end{figure}

Consider now the classical approximation ${\cal C}^0$ to the collision
term, in which one keeps only the cubic terms in the distribution
function. The time evolution of the density of condensed particles is
shown in the figure \ref{fig:nc-C0}, for the same initial condition
and the same values of the UV cutoff.
In this approximation, $n_c(t)$ behaves in a very similar way as with
the unapproximated collision term: the BEC transition happens at the
same time, and the asymptotic density of condensed particles is
approximately the same\footnote{This aspect of the comparison is a bit
  misleading, and is true here only because we have used a strongly
  overoccupied initial condition, for which a large fraction of the
  total number of particles condense in the zero mode. Indeed, the
  classical approximation ${\cal C}^0$, that always tends to put all
  the particles in the condensate when $\Lambda_{_{\rm UV}}\to\infty$
  (see the previous section), is bound to give a similar $n_c$ in such
  a situation.}.

\begin{figure}[htbp]
\begin{center}
\resizebox*{8cm}{!}{\includegraphics{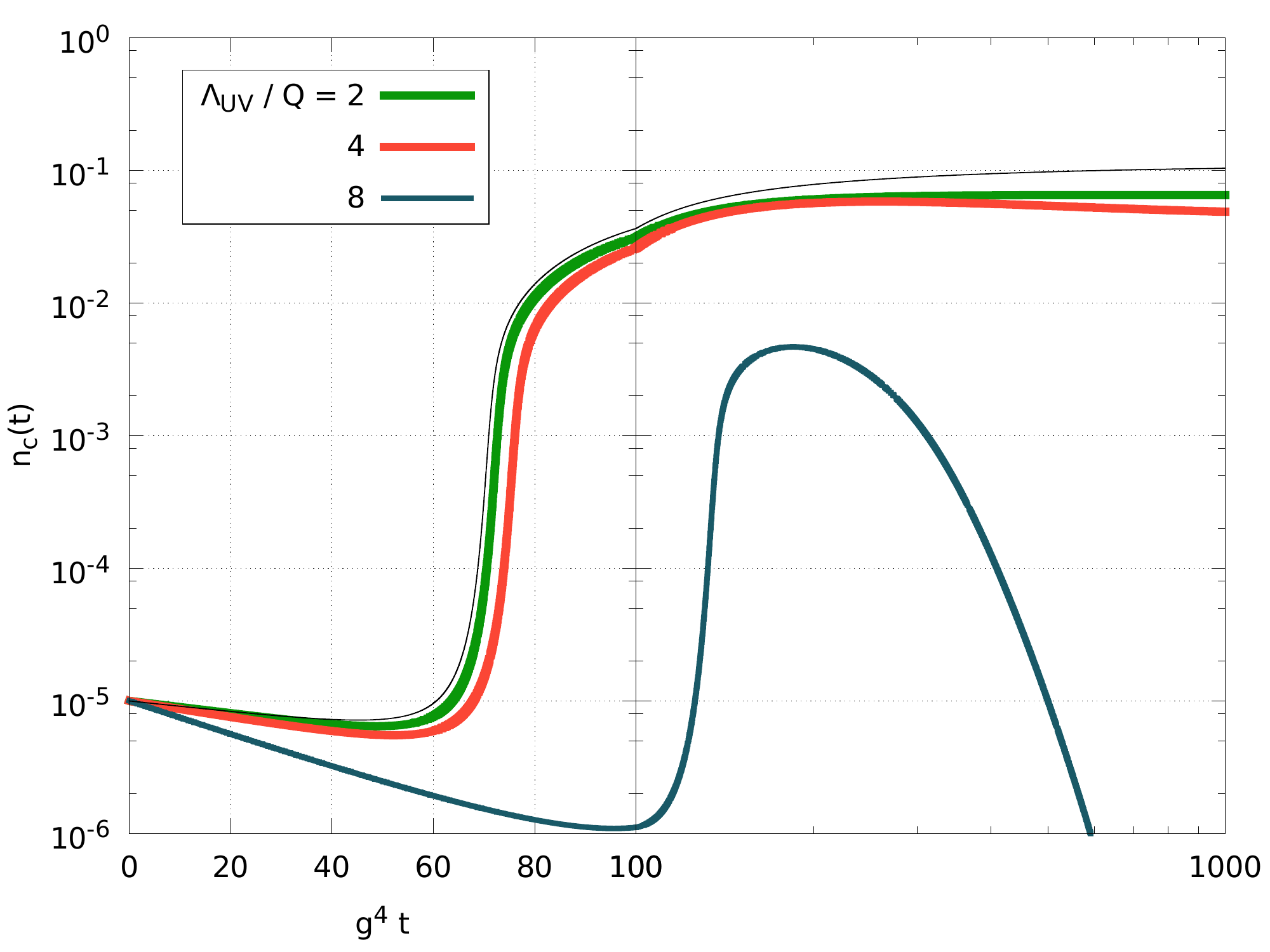}}
\end{center}
\caption{\label{fig:nc-C}Time evolution of the density $n_c$ of
  condensed particles with the collision term in the classical
  approximation ${\cal C}^1$, for three values $\Lambda_{_{\rm
      UV}}/Q=2,4$ and $8$ of the UV cutoff. The thin black line is the
  solution of the unapproximated Boltzmann equation with
  $\Lambda_{_{\rm UV}}/Q=8$.}
\end{figure}

In contrast, the classical approximation ${\cal C}^1$ (see the figure
\ref{fig:nc-C}) tends to underestimate the density of particles in the
condensate. When the UV cutoff is moderately above the physical scale
($\Lambda_{_{\rm UV}}/Q=2,4$ in the figure), the behavior of $n_c(t)$
in this approximation remains in qualitatively good agreement with the
exact result (thin black line). However, for $\Lambda_{_{\rm
    UV}}/Q=8$, the condensate is completely depleted when time goes to
infinity. This is a more dynamical view of what we had learned in the
previous section from the study of conservation laws. At large
$\Lambda_{_{\rm UV}}/Q$, the equilibrium chemical potential is always
negative in this approximation and $n_c(t=\infty)=0$.

\subsection{Evolution of the distribution function}
In the previous subsection, we have seen that the initial condition
(\ref{eq:step}) with $f_0=10$ leads to the formation of a
Bose-Einstein condensate, at a time $g^4 t \approx 70$. If the UV
cutoff is not too large compared to the physical scale $Q$, both
classical approximations reproduce qualitatively this behavior: the
condensation time is almost the same, but the density of condensed
particles differs a bit from the exact value (overestimated in the
approximation ${\cal C}^0$ and underestimated in the approximation
${\cal C}^1$).  Let us now continue this comparison by looking at the
particle distribution itself, at various times.  We do this for a
moderate value of the cutoff, $\Lambda_{_{\rm UV}}/Q=4$, for which the
study of $n_c(t)$ indicated a qualitative agreement of the classical
approximations with the exact quantum result.

\begin{figure}[htbp]
\begin{center}
\resizebox*{8cm}{!}{\includegraphics{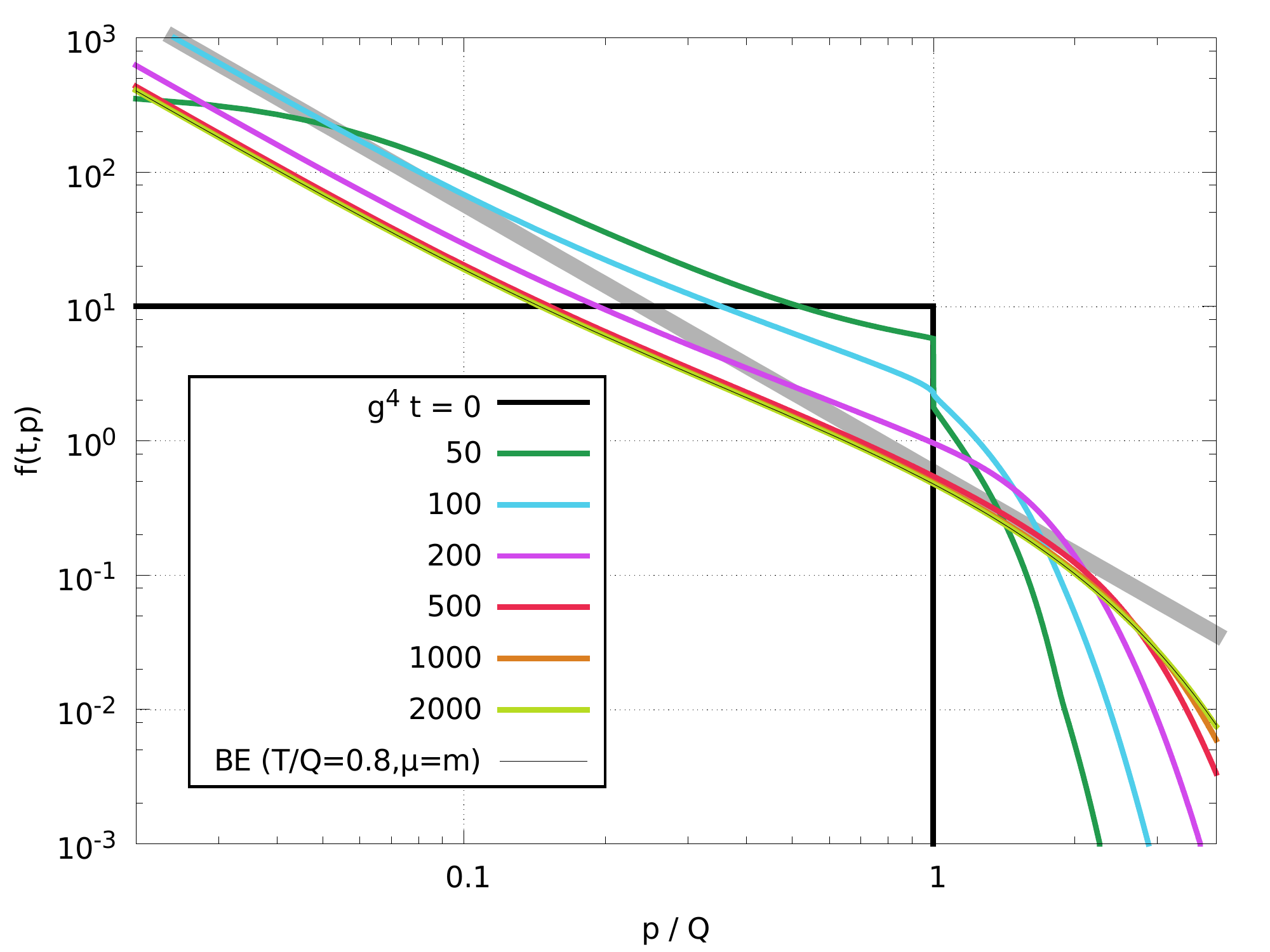}}
\end{center}
\caption{\label{fig:fk-Q}Particle distribution at various times,
  obtained with the unapproximated collision term. The UV cutoff is
  $\Lambda_{_{\rm UV}}/Q=4$. The thin line is a fit with a
  Bose-Einstein distribution with chemical potential $\mu=m$. The gray
  band is a $p^{-2}$ power law.}
\end{figure}

We first show in the figure \ref{fig:fk-Q} the solution of the
Boltzmann equation with an unapproximated collision term. Starting
from a step-like distribution of the form (\ref{eq:step}), the
distribution grows quickly at soft momenta, and also extends to
momenta above $Q$. At times $g^4t \lesssim 70$, i.e. before the BEC
transition, the distribution seems to remain bounded as $p/Q\to 0$.
This soft behavior is drastically changed after the transition, since
we now observe that $f(p)\sim p^{-2}$ as $p/Q\to 0$.  In the soft
sector, the distribution is well fitted by a functional form
$T/(\omega_\p-m)$, which is indeed what one expects in the presence of
a BEC, with a chemical potential equal to the mass of the
particles. Already at $g^4 t =100$, the soft sector of the
distribution is already equilibrium-like, but with a temperature which
is still too high compared to the final equilibrium temperature. The
subsequent evolution is a slow decrease of this temperature,
concomitant with a gradual extension of the tail of the distribution
to larger momenta.

\begin{figure}[htbp]
\begin{center}
\resizebox*{8cm}{!}{\includegraphics{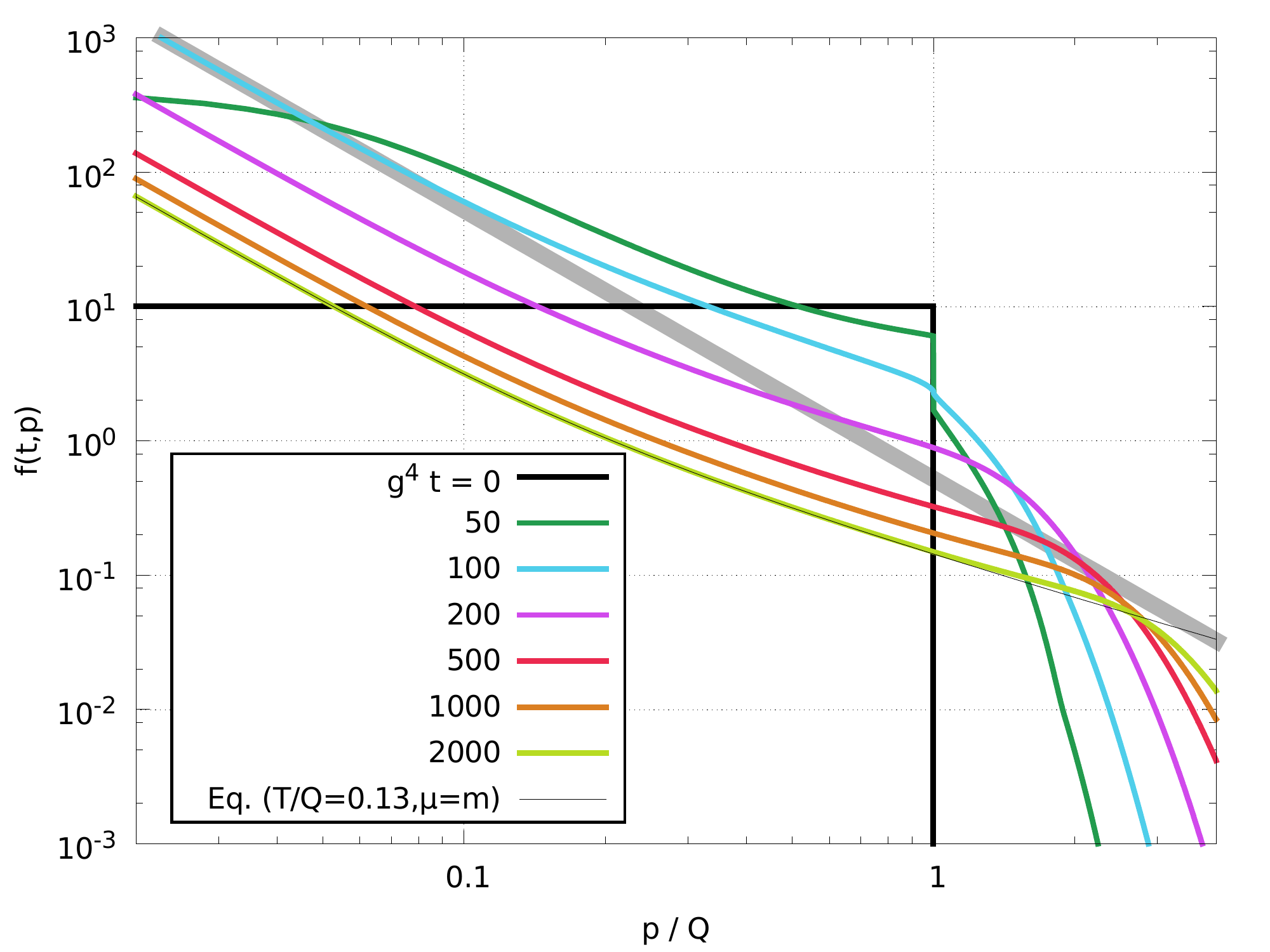}}
\end{center}
\caption{\label{fig:fk-C0}Particle distribution at various times,
  obtained with the collision term in the classical approximation
  ${\cal C}^0$. The UV cutoff is $\Lambda_{_{\rm UV}}/Q=4$. The thin
  line is a fit with a classical equilibrium distribution of the form
  (\ref{eq:class-eq0}) with chemical potential $\mu=m$. The gray band
  is a $p^{-2}$ power law. }
\end{figure}

\begin{figure}[htbp]
\begin{center}
\resizebox*{8cm}{!}{\includegraphics{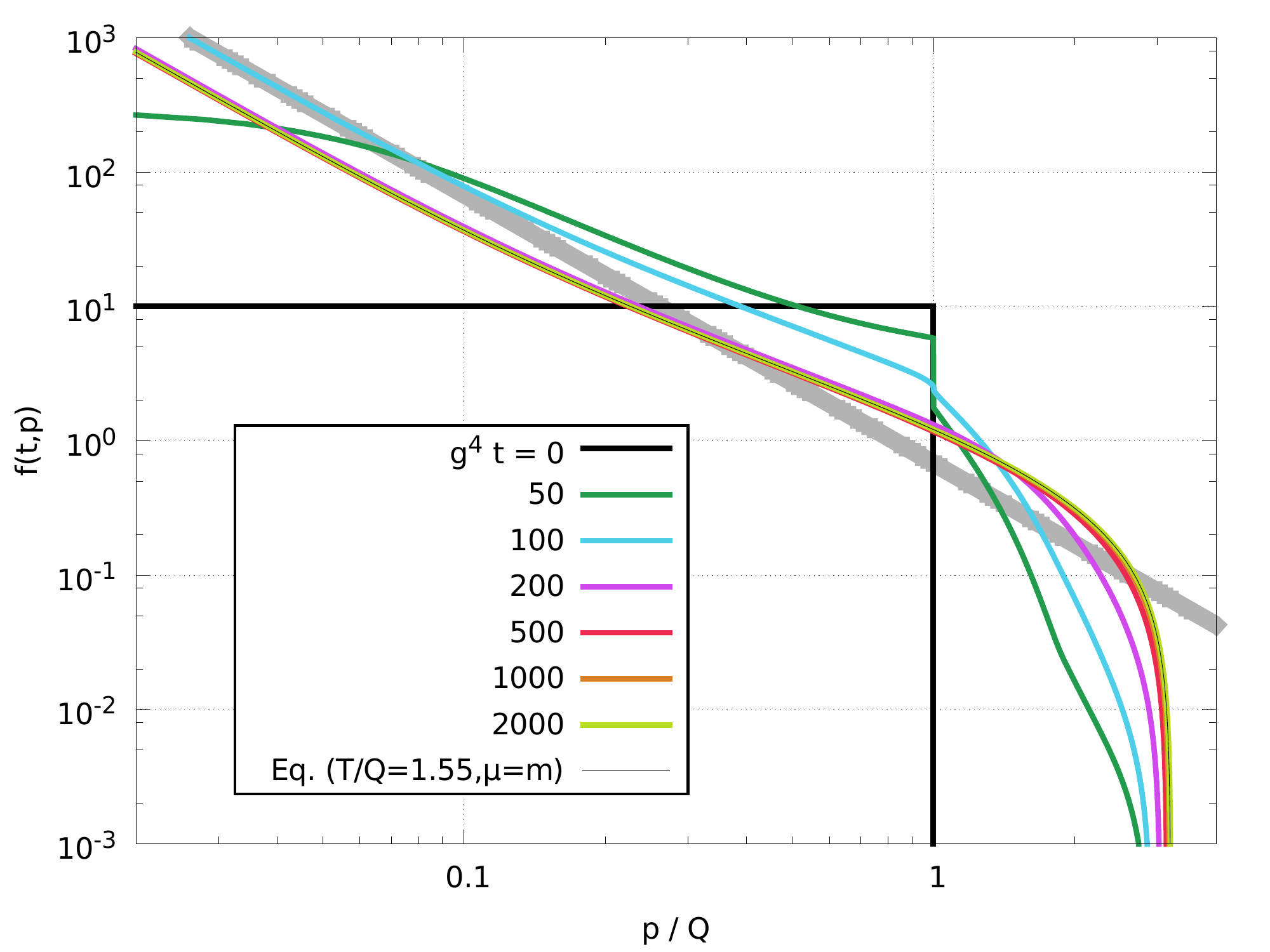}}
\end{center}
\caption{\label{fig:fk-C}Particle distribution at various times,
  obtained with the collision term in the classical approximation
  ${\cal C}^1$. The UV cutoff is $\Lambda_{_{\rm UV}}/Q=4$. The thin
  line is a fit with a classical equilibrium distribution of the form
  (\ref{eq:class-eq1}) with chemical potential $\mu=m$. The gray band
  is a $p^{-2}$ power law. }
\end{figure}

In the figures \ref{fig:fk-C0} and \ref{fig:fk-C}, we show the values
of $f(t,p)$ at the same times, respectively for the classical
approximations ${\cal C}^0$ and ${\cal C}^1$. The first general remark
is that the time-line of events is qualitatively the same as what we
have described above in the quantum case. In fact, up to the BEC
transition and shortly afterwards, the three evolutions are remarkably
close, and the differences become more pronounced only at later times.
The most obvious of these differences lie in the large momentum tail
of the distribution, that behave very differently in the three cases
considered (exact, approximations ${\cal C}^0$ and ${\cal C}^1$). This
is of course expected, since the only common part of the collision
term in these three situations is the term in $f^3$, which is not the
leading one in the tail. Note also that these differences in the
behavior of the tail are amplified when computing the total particle
density or the energy density, since in these calculations the
particle distribution is weighted by $p^2$ and $p^3$
respectively.

In the soft sector, the three computations show a behavior of the form
$T/(\omega_\p-m)$ after the BEC transition, but the values of the
effective temperature $T$ are somewhat different (smaller in the
approximation ${\cal C}^0$ and larger in the approximation ${\cal
  C}^1$).  Moreover, in the approximation ${\cal C}^0$, this effective
temperature needs more time to settle to its equilibrium value.

\section{Conclusions}
\label{sec:concl}
In this paper, we have explored various aspects of the classical
approximation applied to the elastic collision term of the Boltzmann
equation. The starting point of this study is the realization that the
dependence of the effective temperature and chemical potential on the
ultraviolet cutoff in classical statistical field simulations can be
understood in a very simple way from kinetic theory arguments based on
conservation laws. It therefore seemed natural to study more
thoroughly the classical approximation of the Boltzmann equation, and
in particular its sensitivity to the ultraviolet cutoff, as an
alternative to the much more computationally demanding classical
statistical simulations.

Our study is limited to the $2\to 2$ collision term for scalar
particles with a $\phi^4$ interaction.  We have considered two
versions of the classical. In the first one (denoted ${\cal C}^0$),
one keeps only the cubic terms in the distribution function in the
collision term. The CSA analogue of this approximation would be an
initial ensemble of fields whose momentum spectrum falls like
$\sqrt{f(0,\p)}$. The second version of the classical approximation
(denoted ${\cal C}^1$) amounts to replacing $f\to f+1/2$ in the
collision term of the approximation ${\cal C}^0$. By doing this, one
recovers the correct $f^2$ terms, but this also introduces some
spurious terms that are linear in $f$. In the CSA, this approximation
is analogous to an initial ensemble whose spectrum would include
vacuum fluctuations, and thus behaves as $\sqrt{f(0,\p)+1/2}$. 

In the second version of the classical approximation, the collision
term contains terms that are quadratic in the ultraviolet cutoff, even
if the particle distribution has a compact support. These terms, that
are closely related to some non-renormalizable contributions that were
found in the CSA, make the solution of the Boltzmann approximation
sensitive to the UV cutoff even at early times if the cutoff is much
larger than the physical scales.

Without actually solving the Boltzmann equation, one can determine how
its solution behaves at late times, solely from considerations based
on conservation laws. The H-theorem can be generalized to the two
classical approximations considered in this paper, provided that one
modifies a bit the definition of entropy. By solving the conservation
equations, we determined the cutoff dependence of the temperature and
chemical potential that characterize the fixed point.  In both
approximations, these parameters are strongly cutoff dependent and
become very different from the fixed point of the unapproximated
Boltzmann equation if the cutoff is large.

Finally, we have solved numerically the Boltzmann equation, without
any approximation and in the two versions of the classical
approximation, for various values of the ultraviolet cutoff. As
expected, the solutions of the unapproximated Boltzmann equation do
not depend on the UV cutoff as soon as it is a few times larger than
the physical scales. The situation is quite different in the classical
approximation. The approximation ${\cal C}^0$ always leads to the
formation of a Bose-Einstein condensate at large cutoff, regardless of
whether the initial distribution is truly overoccupied or not.  The
opposite happens with the approximation ${\cal C}^1$, where no
condensation takes place at large cutoff. However, if one keeps the
cutoff a few times above the physical scales, it appeared that the two
classical approximations reproduce, at least qualitatively, all the
features of the exact solution. Given the fact that kinetic theory can
account for the cutoff dependence of classical statistical
computations, this analysis conversely suggests that CSA calculations
done with a cutoff not exceeding 4-5 times the physical scales should
be in good qualitative agreement with the unapproximated
underlying quantum field theory.

\section*{Acknowledgements}
We would like to thank J.-P. Blaizot, J. Liao and L. McLerran for
useful discussions about this work.  This work is support by the
Agence Nationale de la Recherche project 11-BS04-015-01.
N.T. is supported by JSPS Strategic Young Researcher Overseas 
Visits Program for Accelerating Brain Circulation (No. R2411).

\appendix

\section{${\bs\Sigma}_{11}$ and ${\bs\Sigma}_{12}$ at two loops (massless case)}
\label{app:sigma}
The collision term can be expressed in terms of the self-energies
${\bs\Sigma}_{11}$ and ${\bs\Sigma}_{12}$. For the $2\to 2$
contribution, we need these self-energies at two loops. Since in the
classical statistical approximation we neglect the vertex
$\Gamma_{1112}$, they are given by a single Feynman diagram,
\setbox1\hbox to 3.5cm{\resizebox*{3.5cm}{!}{\includegraphics{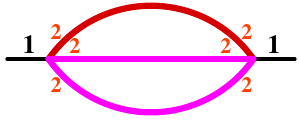}}}
\setbox2\hbox to 3.5cm{\resizebox*{3.5cm}{!}{\includegraphics{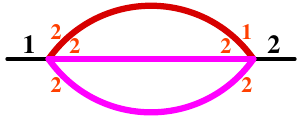}}}
\begin{eqnarray}
-i\big[{\bs\Sigma}_{11}(P)\big]_{_{\rm CSA}}^{\rm 2\ loop}
&=&
\raise -6.5mm\box1
\nonumber\\
-i\big[{\bs\Sigma}_{12}(P)\big]_{_{\rm CSA}}^{\rm 2\ loop}
&=&
\raise -6.5mm\box2\; .
\label{eq:Sigma-CSA}
\end{eqnarray}
As we shall see, these graphs are ultraviolet divergent. We therefore
regularize the integrals by limiting the 3-momentum carried by the
internal $G_{22}$ propagators,
\begin{equation}
|\p|\le \Lambda_{_{\rm UV}}\; .
\end{equation}

The graphs in eqs.~(\ref{eq:Sigma-CSA}) contain as a subgraph the
1-loop $\Gamma_{1122}$ 4-point function (highlighted in purple). In
the massless case, we can use directly the result derived in the
appendix B of ref.~\cite{EpelbGW1}, \setbox2\hbox to
1.1cm{\resizebox*{1.1cm}{!}{\includegraphics{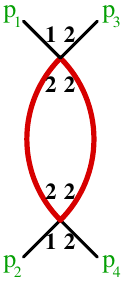}}}
\begin{eqnarray}
  &&{\ }\nonumber\\
  &&{\ }\nonumber\\
  &&
  \smash{\raise -11mm\box2}\equiv\Gamma_{1122}^{\rm 1-loop}(P)=
\nonumber\\
  &&{\ }\nonumber\\
&&
\!\!\!\!\!\!
=-\frac{g^4}{32\pi p}\!\times\!\left\{
\begin{aligned}
&\left[\Lambda_{_{\rm UV}}-\frac{p+|p_0|}{2}\right]
\theta\left(\Lambda_{_{\rm UV}}-\frac{p+|p_0|}{2}\right) &{\scriptstyle[P^2<0]}\\
& \frac{p}{2}
&{\scriptstyle[P^2>0, \frac{p+|p_0|}{2}\le\Lambda_{_{\rm UV}}]}\\
&\Lambda_{_{\rm UV}}-\frac{|p_0|}{2}
&{\scriptstyle [P^2>0, \frac{|p_0|}{2}\le\Lambda_{_{\rm UV}}\le\frac{p+|p_0|}{2}]}\\
& 0
&{\scriptstyle[P^2>0, \Lambda_{_{\rm UV}}\le\frac{|p_0|}{2}]}
\end{aligned}
\right.
\nonumber\\
&&
\label{eq:G1122}
\end{eqnarray}
where $P^\mu\equiv p_1^\mu+p_2^\mu$ is the t-channel momentum.

In terms of this subgraph, the expression for the required
self-energies is
\begin{eqnarray}
-i\big[{\bs\Sigma}_{11}(P)\big]_{_{\rm CSA}}^{\rm 2\ loop}
&=&
\frac{g^4}{3}
\int \frac{d^4K}{(2\pi)^4}\;\pi\delta(K^2)\;\Gamma_{1122}^{1-loop}(P+K)
\nonumber\\
-i\big[{\bs\Sigma}_{12}(P)-{\bs\Sigma}_{21}(P)\big]_{_{\rm CSA}}^{\rm 2\ loop}
&=&
g^4
\int \frac{d^4K}{(2\pi)^4}\;\pi\,{\rm sign}(k^0)\,\delta(K^2)\;\Gamma_{1122}^{1-loop}(P+K)\; .
\nonumber\\
&&
\end{eqnarray}
The two integrals are almost identical, except for the extra factor
${\rm sign}(k^0)$ in the second one.  We can use the delta function
$\delta(K^2)$ in order to perform the integral over $k^0$. The remain
integrals over $k=|\k|$ and $\cos\theta$ (where $\theta$ is the angle
between $\k$ and the external momentum $\p$) are elementary but
require that one carefully splits the integration domain according to
the various cases of eq.~(\ref{eq:G1122}). For $P^2=0$ and $p^0\ge 0$,
this leads to eqs.~(\ref{eq:2loopS0}) and (\ref{eq:2loopS}).

These CSA results contain ultraviolet divergences. However, note that
$\Sigma_{11}$ at two loops is finite if calculated in the full
theory~: \setbox1\hbox to
2.5cm{\resizebox*{2.5cm}{!}{\includegraphics{S11-2loop}}}
\setbox2\hbox to
2.5cm{\resizebox*{2.5cm}{!}{\includegraphics{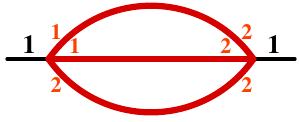}}}
\setbox3\hbox to
2.5cm{\resizebox*{2.5cm}{!}{\includegraphics{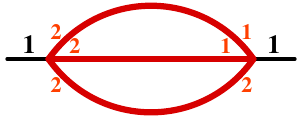}}}
\begin{equation}
-i\big[{\bs\Sigma}_{11}(P)\big]_{\rm full}^{\rm 2\ loop}
=
\raise -4.5mm\box1
+
\raise -4.5mm\box2
+
\raise -4.5mm\box3
\; .
\end{equation}
After some massaging of the integrands, the sum of these three graphs
can be written as
\begin{equation}
-i\big[{\bs\Sigma}_{11}(P)\big]_{\rm full}^{\rm 2\ loop}
=
-\frac{g^4}{12}
\sum_{\epsilon=\pm}
\left[\prod_{i=1}^3\int
\frac{\rmd^4k_i}{(2\pi)^4}
\,\theta(\epsilon k_i^0)2\pi\delta(k_i^2)\right]
(2\pi)^4\delta(\sum_ik_i-P)\; ,
\end{equation}
and it is easy to see that the two terms $\epsilon=\pm$ are separately
ultraviolet finite. The finiteness of $\Sigma_{11}$ at two loops was
expected from the renormalizability of the full
theory\footnote{Intuitively, one should view $\Sigma_{11}$ as a sum of
  cut diagrams, in which the energy can flow only in one direction
  through the cut. At two loops, all the internal lines must be cut,
  so there cannot be any ultraviolet divergent subdiagram.}.  One can
check that ${\bs\Sigma}_{12}$ is also UV finite at two loops in the
full theory.

\section{Numerical evaluation of the collision term}
\label{app:coll}
\subsection{Reduction to a 2-dimensional integral}
In the general case, the collision term is given by a 5-dimensional
integral (three 3-dimensional phase-space integrals, minus 4
constraints coming from the conservation of energy and momentum). This
is very costly to evaluate in a numerical study of the Boltzmann
equation (this 5-dimensional integral would have to be calculated for
every $\p$, at each timestep).

In order to simplify the numerics, we assume that the distribution is
isotropic in momentum space. The scalar interactions being independent
of the direction of the particles involved in the scattering, this
allows to separate completely the angular integrations from the
integrations over the energies~\cite{SemikT1,SemikT2}.  We will
illustrate this in the case of the unapproximated collision term
(\ref{eq:Cfull}), but the same procedure is applicable to the
classical approximations considered in this paper.

\begin{figure}[htbp]
\begin{center}
\resizebox*{8cm}{!}{\includegraphics{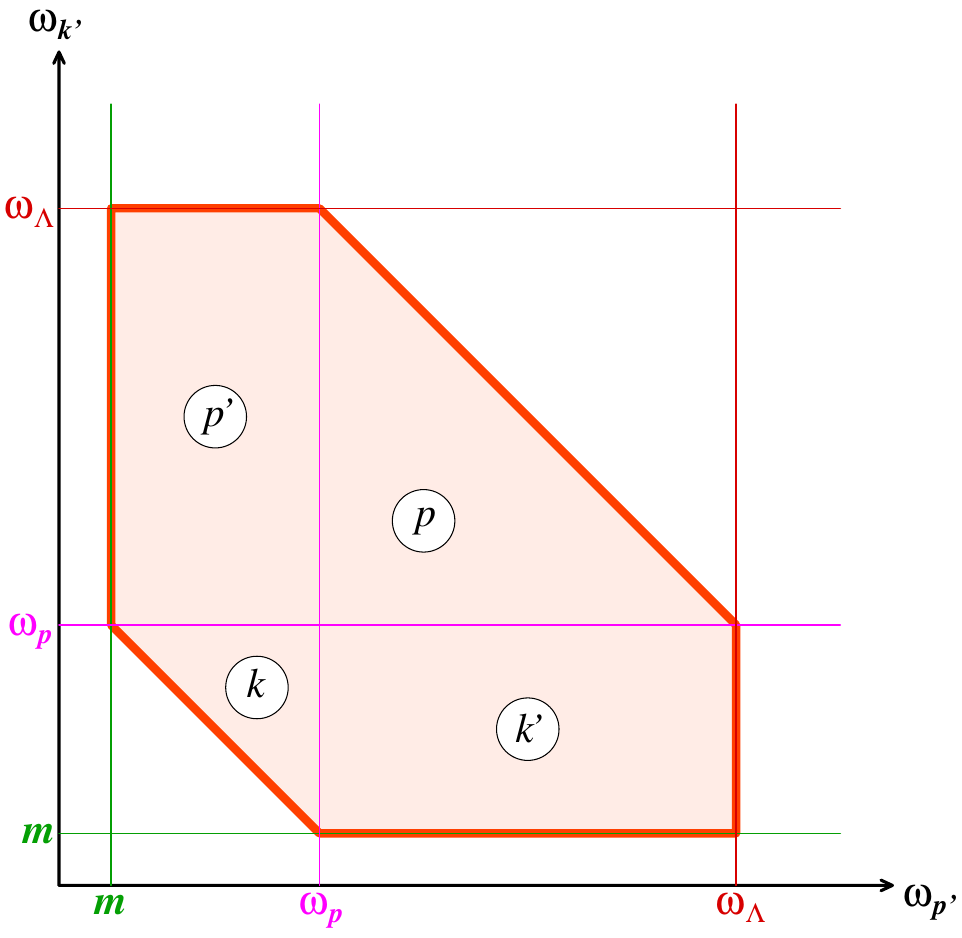}}
\end{center}
\caption{\label{fig:int}Allowed integration domain over the
  variables $\omega_{\p'}$ and $\omega_{\k'}$. The diagonal borders of
  this domain correspond to $\omega_{\k}=m$ and
  $\omega_{\k}=\omega_\Lambda$, respectively. The labels inside the
  white disks indicate the value of ${\rm Min}(pp'kk')$ in the
  corresponding subdomain.}
\end{figure}

Let us call $d\Omega_\p$ the angular measure for the momentum $\p$,
\begin{equation}
d\Omega_\p=\sin\theta_\p\;d\theta_\p\;d\phi_\p\quad,\quad \int d\Omega_\p=4\pi\;,
\end{equation}
where $\phi_\p$ is the azimuthal angle and $\theta_\p$ the polar
angle. Firstly, we can exploit the isotropy in order to write
\begin{eqnarray}
C_\p[f] &=& \frac{1}{4\pi}\int d\Omega_\p\; C_\p[f]\; .
\end{eqnarray}
The second trick is to write the delta function of momentum
conservation as follows,
\begin{equation}
(2\pi)^3\delta(\p+\k-\p'-\k')
=
\int d^3\x\; e^{i(\p+\k-\p'-\k')\cdot\x}\; ,
\end{equation}
which allows to disentangle the angular integrations for the four
momenta. Taking the direction of $\x$ as the polar axis, these angular
integrals  give
\begin{eqnarray}
C_\p[f]&=&\frac{g^4}{32\pi^4 p\omega_\p}\int dp'dkdk'\;\frac{p'kk'}{\omega_{\p'}\omega_\k\omega_{\k'}}\;
\delta(\omega_\p+\omega_\k-\omega_{\p'}-\omega_{\k'})\nonumber\\
&&\times
\int_0^\infty \frac{dx}{x^2}\;\sin(px)\sin(kx)\sin(p'x)\sin(k'x)\;\Big\{f\cdots\Big\}\; ,
\end{eqnarray}
where $\{f\cdots\}$ denotes the combination of $f$'s that enter in the
collision term (only this factor needs to be modified if one wants to
perform the same reduction in the case of the classical
approximations). Using
\begin{equation}
\int_0^\infty \frac{dx}{x^2}\;\sin(px)\sin(kx)\sin(p'x)\sin(k'x)
=
\frac{\pi}{4}\;{\rm Min}(pp'kk')\; ,
\end{equation}
we finally obtain\footnote{We have used $pdp=\omega d\omega$.}
\begin{eqnarray}
C_\p[f]=\frac{g^4}{128\pi^3 p\omega_\p}\int d\omega_{\p'} d\omega_{\k'}\;
{\rm Min}(pp'kk')\;\Big\{f\cdots\Big\}\; ,
\label{eq:coll-2d}
\end{eqnarray}
where the integration over the energy $\omega_{\k}$ has been
eliminated thanks to the delta function. This form of the collision
term is simple enough to be evaluated with high accuracy in a
numerical resolution of the Boltzmann equation. In the figure
\ref{fig:int}, we have represented the allowed integration domain over
the energies $\omega_{p'}$ and $\omega_{\k'}$, taking into account the
fact that all the energies must be between $m$ and
$\omega_{\Lambda}\equiv\sqrt{\Lambda_{_{\rm UV}}^2+m^2}$.
The domain can be divided into four subdomains, according to the value
of ${\rm Min}(pp'kk')$, that we have indicated inside the white disks.

\subsection{Number and energy conserving discretization}
In order to evaluate eq.~(\ref{eq:coll-2d}), we must discretize
momentum space, so that the 2-dimensional integral is replaced by a
double discrete sum. For generic discretizations, the values of $p'$
(obtained from $p,k,k'$ by using energy conservation) will in general
not fall on the grid points, and the value of $f(p')$ must therefore
be obtained by interpolation, which may lead to some errors in the
conservation laws.

A better discretization is to use linearly spaced points from $m$ to
$\omega_\Lambda$ in the energy variables $\omega_\k$ and
$\omega_{\k'}$. If $\omega_\p$ is itself one of these points, then one
is guaranteed that $\omega_{\p'}$ will also be on this grid,
thereby eliminating the need for an interpolation. Let us denote these
grid points $\omega_0=m,\omega_1,\cdots,\omega_{_N}\equiv
\omega_\Lambda$,
\begin{equation}
\omega_i=m+i\Delta\quad,\quad \Delta\equiv \frac{\omega_\Lambda-m}{N}\; .
\end{equation}
We replace an integral over $\omega$ by the following quadrature
formula\footnote{We exclude the index $i=0$ from the sum, because it
  corresponds to $p=0$ where $f(p)$ may be singular. This point is
  handled by using the coupled equations for $f(p)$ and $n_c$
  described in the section \ref{sec:bec}, whose right hand side can be
  similarly reduced to 2-dimensional integrals.}
\begin{equation}
\int_m^\Lambda d\omega\;f(\omega)\quad\to\quad
\Delta\sum_{i=1}^N w_i\;f(\omega_i)\; ,
\end{equation}
where the coefficients $w_i$ depend on the details of the quadrature
formula.

In addition to avoiding the errors due to interpolations, the
conservation of particle number and energy requires that the
discretization exactly preserves the antisymmetry of the integrand if
one swaps initial and final states, and its symmetry if one swaps the
two particles of the initial state, or the two particles of the final
state. In order to achieve this, it is useful to rewrite the
2-dimensional integral that enters in the collision term as follows
\begin{equation}
\int d\omega_{\p'}d\omega_{\k'}\;\cdots
=
\int d\omega_{\p'}d\omega_{\k'}d\omega_\k\;\delta(\omega_\p+\omega_\k-\omega_{\p'}-\omega_{\k'})\;\cdots
\end{equation}
After discretization, if $i$ is the discrete index corresponding to
the energy $\omega_\p$, this integral becomes
\begin{equation}
\Delta^2
\sum_{j,k,l=1}^N
w_j w_k w_l\;\delta_{i+j-k-l}\;\cdots
=
\Delta^2\sum_{k,l=1}^Nw_k w_l w_{k+l-i}\;\cdots
\label{eq:2-sum}
\end{equation}
One can read from this formula the quadrature weights that should
be used in the double sum that appears in the collision term, in
order to preserve exactly\footnote{In practice, due to rounding
  errors, one finds that the particle number and the energy are
  conserved with machine accuracy.} all the symmetries that are
necessary for the conservation laws. 
\begin{figure}[htbp]
\begin{center}
\resizebox*{8cm}{!}{\includegraphics{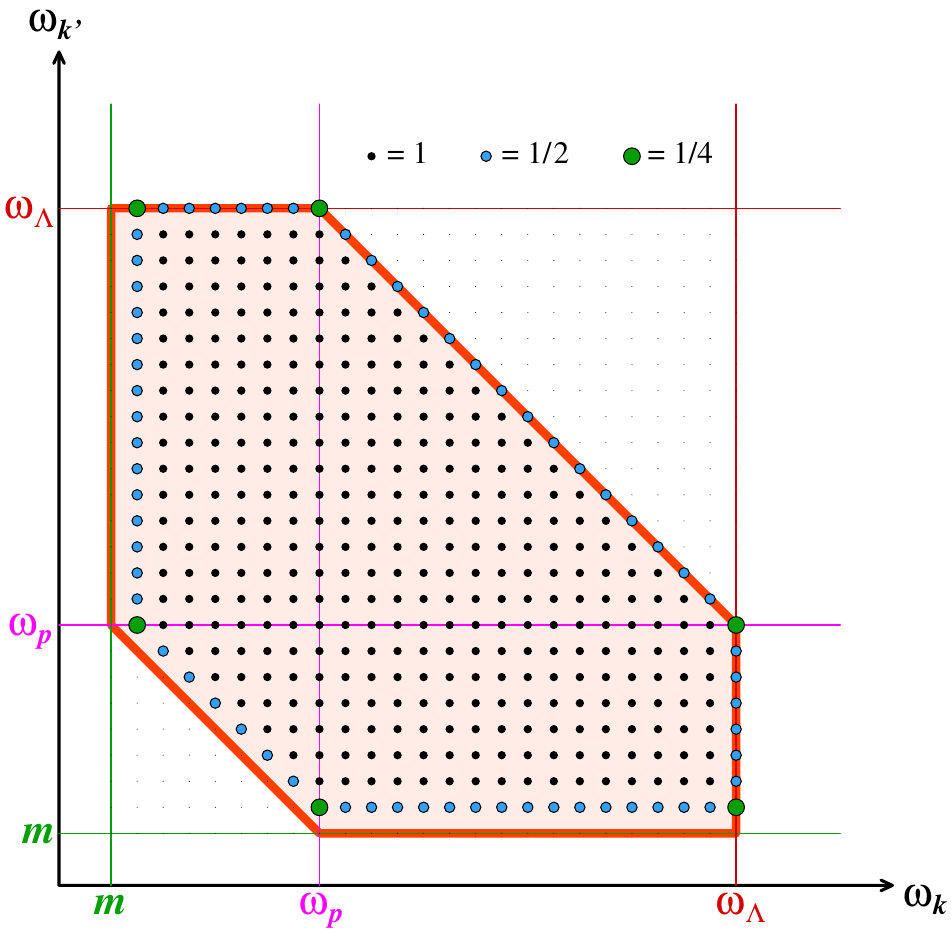}}
\end{center}
\caption{\label{fig:weights}2-dimensional quadrature weights for the
  trapezoidal rule, in the generic case where $\omega_\p$ is in the
  interior of the grid.}
\end{figure}
\begin{figure}[htbp]
\begin{center}
\resizebox*{6cm}{!}{\includegraphics{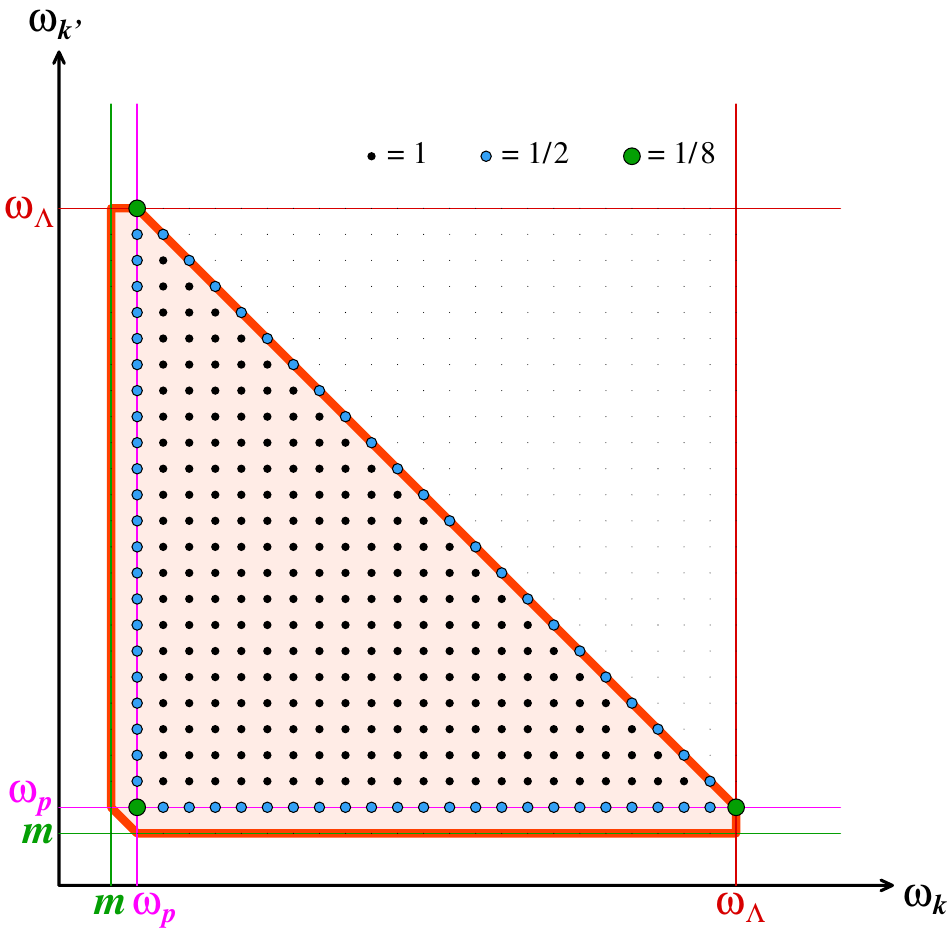}}\hfil
\resizebox*{6cm}{!}{\includegraphics{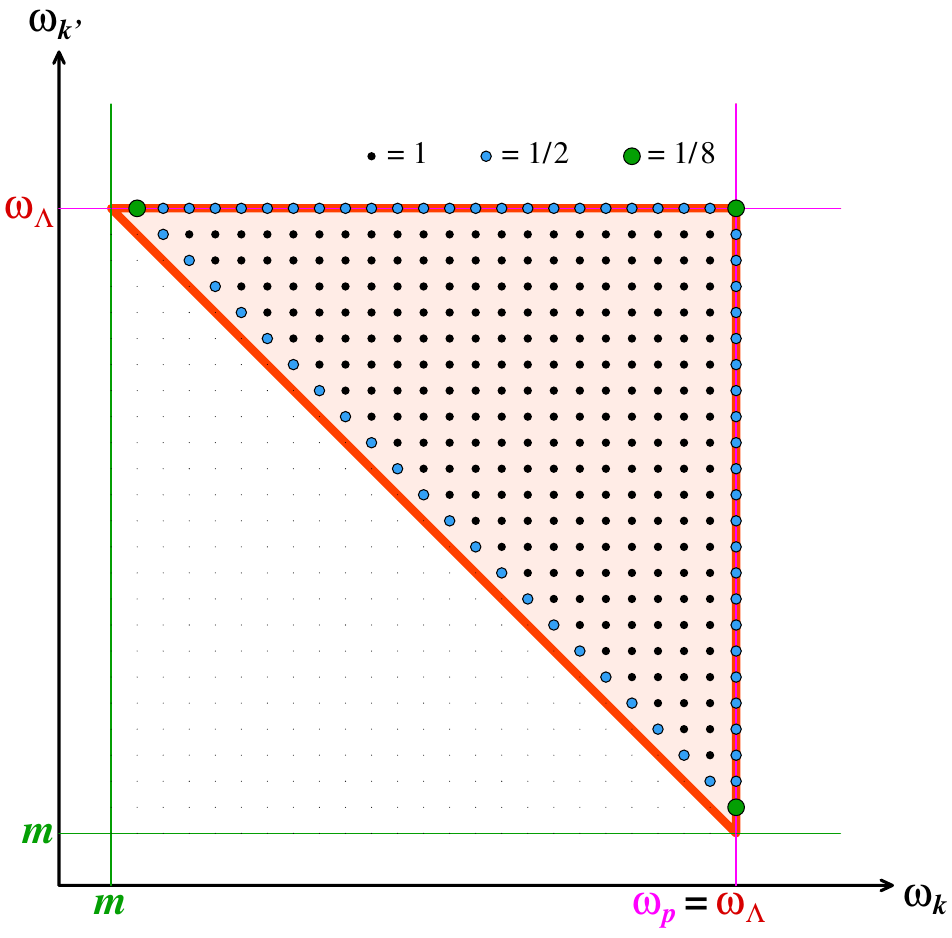}}
\end{center}
\caption{\label{fig:weights_1N}2-dimensional quadrature weights for
  the trapezoidal rule, in the special cases where $\omega_\p$ is at
  one of the endpoints of the grid.}
\end{figure}

The numerical results presented in this paper have been obtained by
using the trapezoidal rule for the 1-dimensional integration~:
\begin{equation}
w_1=\frac{1}{2}\quad,\quad
w_{2,3,\cdots,N-1}=1\quad,\quad
w_{N}=\frac{1}{2}\; .
\end{equation}
For this choice, the 2-dimensional weights that lead to an exact
conservation of particle number and energy are represented in the
figures \ref{fig:weights} (generic case) and \ref{fig:weights_1N}
(when $\omega_\p$ is at one of the endpoints of the grid). Similar
constructions can be done for higher order quadratures,
such as Simpson's rule.

As a final note, let us stress that the computation time of the double
sum in eq.~(\ref{eq:2-sum}) scales as $N^2$. At each time step, this
must be repeated for each of the $N$ values of energy. Therefore, the
total computation time scales as $N^3$ times the number of time
steps. This strong dependence on $N$ puts a practical limit on the
number of discrete values of the energy one can take (in the
calculations presented in this paper, the largest value of $N$ that we
used is $N=8000$.).
\begin{figure}[htbp]
\begin{center}
\resizebox*{8cm}{!}{\includegraphics{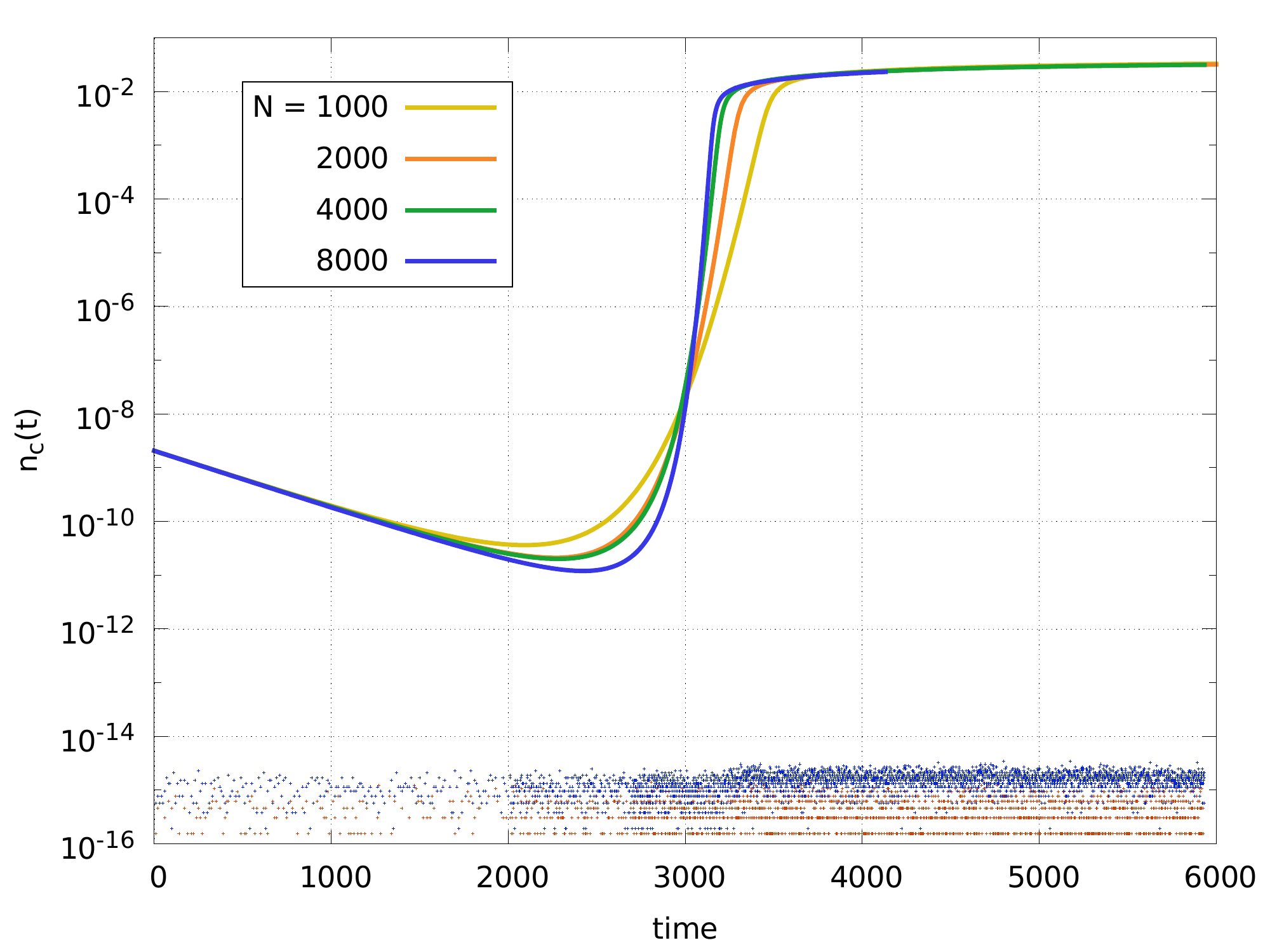}}
\end{center}
\caption{\label{fig:example} Numerical solution of
  eqs.~(\ref{eq:coupled1}) and (\ref{eq:coupled2}) for an overoccupied
  initial condition. The curves show the time evolution of the density
  of condensed particles, $n_c(t)$, for various numbers $N$ of grid
  points. The scattered points at the bottom are the relative errors
  on the conservation of particle number (orange points) and energy
  (blue points).}
\end{figure}

In the figure \ref{fig:example}, we show an example of the evolution
of the density $n_c$ of condensed particles (for an overoccupied
initial condition), obtained by solving the coupled equations
(\ref{eq:coupled1}) and (\ref{eq:coupled2}) with the above
discretization method. The scattered points are the relative error on
the conservation of the number of particles (orange points) and of
energy (blue points). The various curves correspond to different
numbers of grid points in the discretization of the energy axis. Since
we are considering only spatially homogeneous systems, the
condensation transition should be a genuine discontinuity. However, in
any discretization of momentum space, there will be a ``smallest
nonzero momentum'', that in a sense plays the role of an inverse size
of the system. One can see in the figure \ref{fig:example} that by
increasing $N$ (i.e. increasing this system size), the transition seen
in the numerical calculation becomes sharper. Note that in order to
handle this fast transition, it is best to use an adaptative timestep,
adjusted dynamically in order to limit the variation of $n_c$ during
one step.

\section{Linear part of ${C}^{{\cal C}^1}_\p[f]$}
\label{app:lin}
In the classical approximation ${\cal C}^1$, the collision term is the
sum of the full quantum collision term (that contains cubic and
quadratic terms in $f$) and of an extra piece which is linear in
$f$. After a tedious calculation, this additional term can be reduced
in the massless case to a formula that contains at most 1-dimensional
integrals,
\begin{align}
{C}_{\p,{\rm lin}}^{{\cal C}^1}[f]
=\null&\frac{ g^{4}}{512p^2\pi^{3}} \left\{\frac{pf(p)}{6}(2p^2-3\Lambda_{_{\rm UV}}^2)
+\int\limits_{0}^{p}   d k\,k(2\Lambda_{_{\rm UV}} +k)\,f(k)\right.\notag\\
&+ p (2\Lambda_{_{\rm UV}}+p)\int\limits_{p}^{\Lambda_{_{\rm UV}}} d k\,f(k)
+\int\limits_{\Lambda_{_{\rm UV}}-p}^{\Lambda_{_{\rm UV}}} d k\,f(k)(p+ k-\Lambda_{_{\rm UV}})^2\notag\\
&\left.-3p\int\limits_{0}^{\Lambda_{_{\rm UV}}}d k\,k\,f(k)\right\}\;.
\label{eq:Clin}
\end{align}
From this formula, one can check explicitly that this extra term also
conserves particle number and energy, since it satisfies
\begin{eqnarray}
\int_0^{\Lambda_{_{\rm UV}}}dp\;p^2 \;{C}_{\p,{\rm lin}}^{{\cal C}^1}[f] &=&0\; ,\\
\int_0^{\Lambda_{_{\rm UV}}}dp\;p^3\;{C}_{\p,{\rm lin}}^{{\cal C}^1}[f] &=&0\; .
\end{eqnarray}
From eq.~(\ref{eq:Clin}), it is also easy to extract the leading power
in the UV cutoff. If we assume that the support of $f$ does not extend
beyond $\Lambda_{_{\rm UV}}/2$ and that $p<\Lambda_{_{\rm UV}}/2$, we
get
\begin{equation}
{C}_{\p,{\rm lin}}^{{\cal C}^1}[f]
=
-\frac{g^4\Lambda_{_{\rm UV}}^2}{1024\pi^3}\,\frac{1}{p}\;f(\p)+\cdots
\end{equation}

Note also that the linear part of the collision term can be written as
\begin{equation}
{C}_{\p,{\rm lin}}^{{\cal C}^1}[f]
=
\frac{ g^{4}}{512p^2\pi^{3}}
\int_0^{\Lambda_{_{\rm UV}}} dp'\;K(p,p')\;f(p')\; ,
\label{eq:Clin1}
\end{equation}
with
\begin{eqnarray}
K(p.p')
&\equiv&
\frac{p(2p^2-3\Lambda_{_{\rm UV}}^2)}{6}\delta(p-p')
+{\rm Min}(p(2\Lambda_{_{\rm UV}}+p),p'(2\Lambda_{_{\rm UV}}+p'))
\nonumber\\
&&\quad
-3pp'
+\theta(p+p'-\Lambda_{_{\rm UV}})\;(p+p'-\Lambda_{_{\rm UV}})^2\; .
\end{eqnarray}
The kernel $K(p,p')$ is real symmetric, and therefore the linear part
of the collision term, if viewed as a linear operator in the space of
$f(p)$'s, has a real spectrum and its eigenfunctions form an
orthonormal basis. It is easy to check that this linear operator has a
2-dimensional null eigenspace, made of all the distributions of the
form $f(p)=A+B(p/\Lambda_{_{\rm UV}})$. All the other eigenvalues are
negative, which means that if we keep only this linear term in the
Boltzmann equation, the solutions are attracted to this null space.
Note that this fixed point is not realized in practice, because the
terms in $f^3$ and $f^2$ become important before reaching it and
modify the functional form of the allowed fixed points.

\section{Dependence of $n_c$ on the UV cutoff}
\label{app:masslessnc}
If a Bose-Einstein condensate exists in the system, one has $\mu=m$.
As a result, $\wt{T}$ and $\wt{n}_c$ can be solved from
eqs. (\ref{eq:class-eq}), which take the form
\begin{eqnarray}
&&\wt{T}= 6\pi^2(\wt\epsilon- \wt{m}\tilde{n})+ 3 f(\wt{m})-\frac{1}{2}\wt{m}\;,\nonumber\\
&&\wt{n}_c={\wt n} +\frac{1}{12\pi^2}-\frac{\wt T}{2\pi^2}\int_0^1 \rmd{x}\;\frac{x^2}{\sqrt{x^2+{\wt m}^2}-\wt{m}}\;.
\end{eqnarray}
In the massless limit $m\to 0$, one has
\begin{eqnarray}
T=\frac{3}{8}\Lambda_{_{\rm UV}}+\frac{6 \pi ^2 \epsilon}{\Lambda_{_{\rm UV}} ^3}\;,\qquad 
n_c=n-\frac{3 \epsilon}{2 \Lambda_{_{\rm UV}} }-\frac{\Lambda_{_{\rm UV}} ^3}{96 \pi ^2}\;.
\end{eqnarray}
For the initial condition (\ref{eq:step}), the condition for the
existence of a Bose-Einstein condensate is given by
\begin{equation}
\frac{\Lambda_{_{\rm UV}}}{Q}< 2 \left[2f_0\left(1 - \frac{9}{8}\frac{Q}{\Lambda_{_{\rm UV}}}\right)\right]^{\frac{1}{3}}\; .
\end{equation}
For any $f_0$, there is always a limiting value of $\Lambda_{_{\rm
    UV}}$ above which there cannot be a BEC.

Similarly, in the case of the classical approximation $\mathcal{C}^0$ one has
\begin{eqnarray}
T=\frac{6 \pi ^2 \epsilon}{\Lambda_{_{\rm UV}} ^3},\qquad n_c=n-\frac{3 \epsilon}{2 \Lambda_{_{\rm UV}} }\; ,
\end{eqnarray}
and the condition for the existence of a Bose-Einstein condensate is given by
\begin{equation}
\Lambda_{_{\rm UV}}>\frac{3\epsilon}{2n}=\frac{9}{8}Q\; .
\end{equation}
For the approximation ${\cal C}^0$, the situation is reversed: at
large enough cutoff, a BEC would always be present.

%\bibliographystyle{unsrt}
%\bibliography{biblio}

\end{document}